\RequirePackage{fix-cm}
\documentclass{svjour3}                     
\smartqed  

\usepackage{lineno,hyperref}

\usepackage{graphicx}%
\usepackage{multirow}%
\usepackage{amsmath,amssymb,amsfonts}%
\usepackage{mathtools}
\usepackage{mathrsfs}%
\usepackage[title]{appendix}%
\usepackage{xcolor}%
\usepackage{textcomp}%
\usepackage{manyfoot}%
\usepackage{booktabs}%
\usepackage{algorithm}%
\usepackage{algorithmicx}%
\usepackage{algpseudocode}%
\usepackage{listings}%
\usepackage{subcaption}

\usepackage[section]{placeins}
\DeclareMathOperator{\spn}{span}

\usepackage[numbers,sort&compress]{natbib}
\begin{document}

\title{Effects of Internal Resonance and Damping on Koopman Modes
}


\author{Rahul Das         \and
        Anil K Bajaj      \and
        Sayan Gupta
}


\institute{Rahul Das \at
              Department of Applied Mechanics and Biomedical Engineering, Indian Institute of Technology, Madras, India. \\
              \email{rahuldas93.rd11@gmail.com}           
           \and
           Anil K. Bajaj \at
              School of Mechanical Engineering, Purdue University, Purdue, USA.
              \email{bajaj@purdue.edu}
           \and
           Sayan Gupta \at
              Depatment of Applied Mechanics and Biomedical Engineering, Indian Institute of Technology, Madras, India. \at Center for Complex Systems and Dynamics, IIT Madras, India.
              \email{sayan@iitm.ac.in}
}
\date{Received: date / Accepted: date}
%
%
\maketitle
\begin{abstract}
This study investigates the nonlinear normal modes (NNMs) of a system comprising of two coupled Duffing oscillators, with one oscillator being grounded and  with the coupling being both linear and nonlinear. The study utilizes the eigenfunctions of the Koopman operator and validates their connection with the Shaw-Piere invariant manifold framework for NNMs. Furthermore, the study delves into the impact of internal resonance and dissipation on the accuracy of this framework by defining a continuous quantitative measure for internal resonance. The applicability and robustness of the framework for the systems which are very similar qualitatively to that of an ENO, are also observed and discussed about the limitations of the approximation technique.
\end{abstract}
\section{Introduction}\label{sec1} 
In classical dynamical theory \cite{perko2013differential}, the eigensubspaces \cite{hoffmann1971linear,guckenheimer2013nonlinear,MR1129886} of a linear dynamical system are referred to as Linear Normal Modes (LNMs), which are invariant subspaces in terms of flow. LNMs play a crucial role in decoupling the dynamical system and the general solution is obtained by solving its components separately, relying on the principles of linear independence and linear superposition.
However, when applied to nonlinear dynamical systems, LNMs often fall short, failing to capture important phenomena such as bifurcation, chaos, and nonlinear internal resonance \cite{DAS2023104285}. Recognizing these limitations, Lyapunov, Kauderer and Rosenberg \cite{lyapunov1992general,kauderer1958schwingungslehre,10.1115/1.3641668,10.1115/1.3636501,10.1115/1.3643948} pioneered the concept of \textit{Nonlinear Normal Modes} (\textit{NNM}s). NNMs extend LNM theory to nonlinear systems, while preserving the invariance property of normal modes for underlying Hamiltonian systems. Shaw and Pierre \cite{shaw1991non,shaw1993normal} further advanced the concept for dissipative systems by introducing \textit{invariant manifolds} to compute NNMs and investigate their topology. However, Shaw's technique involves solving numerous polynomial equations, posing computational challenges for systems with a high degree of freedom. Moreover, the accuracy of the manifold parameterization diminishes significantly for the states (points in state space) further from the equilibrium point amidst strong nonlinear interactions. This motivated the development of various frameworks to study Shaw-Pierre type NNMs. These include an algebraic geometry-based computational framework \cite{petromichelakis2021computational}, the operator-theoretic approach \cite{cirillo2016spectral}, and the introduction of the \textit{Spectral Sub-Manifolds} (\textit{SSMs}) \cite{haller2016nonlinear,ponsioen2018automated,li2023model}. 
A concise overview of the development and contributions of these articles is presented in Sec.\ref{sec2},  underscoring the motivation to explore NNMs using an operator-theoretic approach.

In contrast to classical dynamical theory, an operator-theoretic approach has been developed by B.O. Koopman \cite{koopman1931hamiltonian}, focusing on the spectral properties of the \textit{Koopman operator}. Cirillo \textit{et. al.} \cite{cirillo2016spectral} developed the framework for studying Shaw-Pierre NNMs (\textit{invariant manifolds}) by adapting the Koopman operator theory. In this method, the parameterization of the manifold is carried out by a nonlinear coordinate transformation, which involves mapping the state variable into an infinite-dimensional vector space. 
A proof of the uniqueness of this mapping is however not available. Furthermore, the works of Sternberg \cite{sternberg1957local,sternberg1957localreal}, Mezi\'{c} \cite{mezic2013analysis} and Lan \cite{lan2013linearization} indicate that this framework is exclusive to systems with only hyperbolic fixed points, and the analytic properties of the nonlinear coordinate transformation are compromised in the case of systems with nonresonating eigenvalues. Cirillo \textit{et. al.} \cite{cirillo2016spectral} have explored the impact of resonating eigenvalues on the accuracy of the Koopman operator framework, particularly in systems with \textit{linear coupling terms}. For such cases, the presence of strong nonlinearity in coupling leads to strong nonlinear internal resonance, even for systems with non-resonating eigenvalues. 

The motivation behind the present study is threefold: first, to provide a proof of the uniqueness of mapping the state vector into the infinite-dimensional vector space (as stated in Proposition \ref{pro1}); second, to investigate the effects of nonlinear internal resonance and damping on the accuracy of this framework; and finally, to explore the applicability and the accuracy of the Koopman operator framework for  systems which are very similar qualitatively to systems with at least one nonhyperbolic eigenvalue.
%
In this study, we have considered a $2$-dof system comprising of two Duffing oscillators, such that the first oscillator is grounded and the second oscillator is coupled to the first one, where the coupling  has additive linear and nonlinear terms. 
The choice of the system offers flexibility to meet the objectives of this study. The presence of nonlinearity in the coupling  ensures the possibility of nonlinear resonance and by making the linear part of the coupling term close to zero, the second oscillator will approach to be \textit{essentially nonlinear} (ENO) (\textit{i.e.} the linear part of the stiffness is zero). The corresponding dynamical system contains at least one non-hyperbolic fixed point.

This paper is structured as follows: Sec.\ref{sec2} is devoted to a literature review of NNM, which motivates the study of Koopman operator framework for determining NNMs. Sec.\ref{sec3} presents  brief discussions on the principle and the definition of the Koopman operator along with its spectral properties. The relation between Koopman eigenfunctions and the LNMs is also established in Sec.\ref{sec3}. Sec.\ref{sec4} presents some theorems on linearization; and a nonlinear coordinate transformation is carried out from the corollary (Corollary \ref{corl1}) of these theorems, which eventually leads to the proof of the relation between Koopman eigenfunctions and NNMs in Sec.\ref{sec5.1}. Sec.\ref{sec5} introduces Proposition \ref{pro1}, which proves the uniqueness of the representation of the state variables in terms of Koopman eigenfunctions; laying the foundation for the parameterization of the invariant manifolds. Sec.\ref{sec6} delves into the solution algorithm for the system being considered in this study (which potentially exhibits nonlinear internal resonance with nonresonant eigenvalues). In Sec.\ref{sec7}, the primary objectives of the work are outlined, emphasizing the need for a parametric study. This section also presents the results obtained from numerical simulations, accompanied by relevant discussions. In addition, an empirical measure is introduced in this section for quantifying internal resonance, which captures the impact of the internal resonance on the Koopman modes in various resonance conditions. At the end of this section, the approximation accuracy at different damping regimes and the applicability of this procedure for an \textit{essentially nonlinear oscillator} are detailed with relevant figures. Finally, Sec.\ref{sec8} summarizes the salient features of this study.
\section{Review on Nonlinear Normal Modes}\label{sec2}
The pioneering contributions of Lyapunov, Kauderer, and Rosenberg \cite{lyapunov1992general,kauderer1958schwingungslehre,10.1115/1.3641668,10.1115/1.3636501} have laid the foundation for the theory of NNMs, extending the invariance principle of LNMs to address the complexity arising from nonlinearities in dynamical systems. Rosenberg \cite{10.1115/1.3641668,10.1115/1.3636501,10.1115/1.3643948} specifically focused on $n$-degree of freedom conservative systems, defining NNMs as vibrations in unison, \textit{i.e.} synchronous periodic motions, where all material points of the system reach their extrema values at the same instant of time and cross the equilibrium simultaneously. These NNMs were categorized into either straight modal lines, known as similar NNMs \cite{10.1115/1.3641668,10.1115/1.3636501} or modal curves for non-similar NNMs \cite{rosenberg1964nonsimilar,VAKAKIS1992341} in the configuration space of the system; see Appendix \ref{secA} for more detailed explanations.
This representation directly originates from the generalization of Linear Normal Modes (LNMs) for nonlinear cases, where the normal modes are depicted as the relationship between the displacement of the material points of the system. This class of representation is commonly referred to as Kauderer-Rosenberg NNMs.

Similar NNMs occur only in special cases, such as in homogeneous and symmetric systems. These NNMs are qualitatively similar to LNMs as the relationship between the displacement of the material points of the system is linear, facilitating straight modal lines in configuration space \cite{mikhlin2010nonlinears}. Conversely, non-similar normal modes represent more general cases, where the relationship between the displacement of the material points of the system is nonlinear, resulting in modal curves in configuration space; see Appendix \ref{secA}. This nonlinear relation makes computation of non-similar NNMs, more challenging compared to similar NNMs \cite{10.1115/1.3629599,rosenberg1964nonsimilar,rand1971higher}. Manevitch and Mikhlin introduced a power series method \cite{manevich1972periodic,mikhlin1996normal}, later developed into \textit{Complexification-Averaging} (\textit{CX-A}) technique by Manevitch \cite{manevitch1999complex,manevitch2001description,manevitch2003}, for computing non-similar NNMs for conservative systems. Nayfeh \cite{nayfeh1994nonlinear,nayfeh1996nonlinear} extended this technique for  continuous systems. Vakakis \cite{vakakis2001,vakakis2003dynamics,vakakis2008nonlinear}, Gendelman \cite{gendelman2001energy,gendelman2008attractors}, Kerschen \cite{kerschen2009nonlinear,kerschen2014modal} and others \cite{starosvetsky2008attractors,starosvetsky2008strongly,viguie2009energy,viguie2009nonlinear,sapsis2009efficiency} conducted detailed investigation of energy transfer mechanism between two substructures of a multi degree of freedom dissipative system, focusing on developing methodologies to compute Kauderer-Rosenberg type NNMs using Manevitch's CX-A technique for various engineering problems. These investigations involve study of the underlying Hamiltonian structure, despite these systems being essentially non-conservative. This makes the notion of NNMs less robust for non-conservative systems.

 The concept of invariant manifolds, based on \textit{centre manifold technique} \cite{carr1981applications,guckenheimer2013nonlinear,glendinning1994stability} has been introduced by Shaw and Pierre \cite{shaw1991non,shaw1993normal,shaw1994normal,pierre2006nonlinear} to define NNMs for non-conservative systems. They defined NNMs as $2$-dimensional invariant manifolds (see Appendix \ref{sec_def}), embedded in $n$-dimensional state space, and are tangent to the LNMs (linear eigenspaces) of the linearized system at the equilibrium points. Shaw's technique requires solving a large number of nonlinear algebraic equations, depending on the desired degree of approximation and the type of nonlinearity in the system; making this approach computationally challenging for systems with higher degrees of freedom. Furthermore, parameterization works only in the vicinity of equilibrium in the presence of strong nonlinear interactions \cite{haller2016nonlinear,ponsioen2018automated,li2023model,cirillo2016spectral}. Petromichelakis \textit{et.al.} \cite{petromichelakis2021computational} developed a computational algebraic geometrical technique based on the concept of Gr\"{o}bner basis to address the computational challenges. More recently, Nield \textit{et. al.} \cite{neild2015use} reported that the Shaw–Pierre type invariant surfaces are non-unique in the linearized system, raising uncertainties about the persistence of the parameterization of the manifolds. Haller \textit{et. al.} \cite{haller2016nonlinear,ponsioen2018automated,li2023model} introduced a generalized definition of NNMs using \textit{Spectral Sub-Manifolds} (\textit{SSM}s), defined as the smoothest member of an invariant manifold family tangent to a modal sub-bundle along a NNM, to address the uniqueness of the Shaw-Pierre type NNMs. Though these methodologies are well developed, the issue with non-global parameterization remains unresolved.

Cirillo \textit{et. al.} \cite{cirillo2016spectral} addressed the challenges associated with the non-uniqueness and limited non-global validity of parameterization in Shaw-Pierre NNMs, by adapting an operator theoretic approach, named \textit{Koopman Operator Theory} \cite{koopman1931hamiltonian}. This approach and related  literature \cite{mezic2013analysis,lan2013linearization} suggest that only one of these invariant manifolds is infinitely many times differentiable ({\it i.e.}, is analytic), which is a necessary condition for unique parameterization of the manifold. 
Without further analysis, it has been stated  that there exists a unique, analytic Shaw-Pierre type invariant surface tangent to any two-dimensional modal subspace of a nonlinear system if the eigenvalues of the linearized system are nonresonant. Lan and Mezi\'c \cite{lan2013linearization} have claimed that global validation of parameterization can be achieved with the Koopman operator framework. However, while this framework accurately predicts the trajectories for nonlinear oscillators with  \textit{ linear coupling terms} (like the systems considered in \cite{cirillo2016spectral}), the presence of any nonlinearity in the coupling  often results in strong nonlinear interactions, inducing nonlinear internal resonance. Recognizing these potential challenges, the motivation for further development of the Koopman operator framework and extension of the domain of application appears to be a necessary and promising direction for further research in NNMs.  
\section{Koopman operator}\label{sec3}
The Koopman operator theory \cite{koopman1931hamiltonian} provides a framework for studying nonlinear dynamical systems by mapping 
the state space variable to an infinite-dimensional vector space, such as the Hilbert space \cite{debnath2005introduction}. The fundamental principle of this framework in the calculation of NNMs, lies in the spectral properties \cite{hoffmann1971linear,MR1129886} of the Koopman operator, with the objective being estimating the eigenfunctions of the operator and representing the dynamics in the Koopman operator spectrum. This is equivalent to decoupling the equations of the nonlinear dynamical system (see Sec.\ref{sec4}) and representing the dynamical response as a linear combination of NNMs. 

\subsection{Definition of the Koopman operator}
\begin{figure}[htbp]%
\centering
\subcaptionbox{\label{fig1a}}{\includegraphics[width=0.4\textwidth]{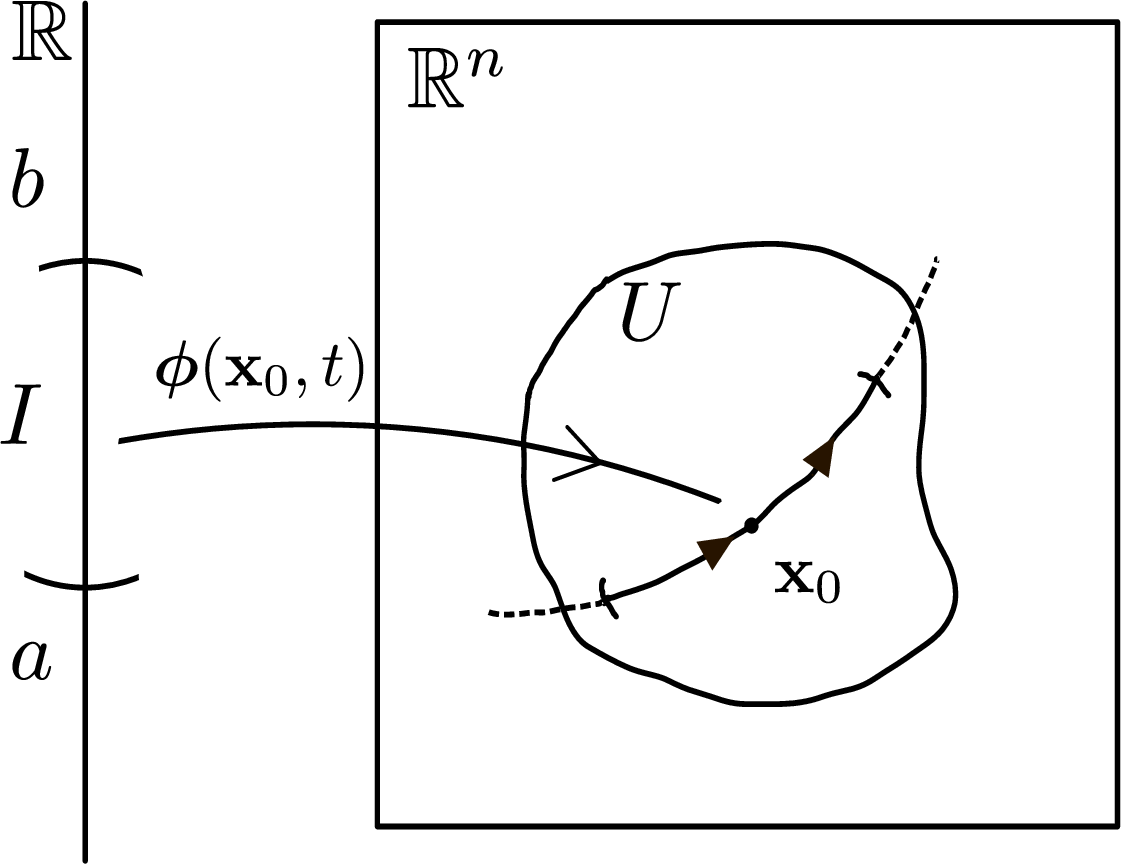}}\hfill
\subcaptionbox{\label{fig1a}}{\includegraphics[width=0.4\textwidth]{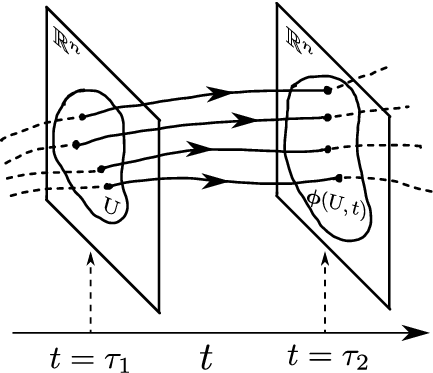}}
\caption{A solution curve and the flow : (a) The solution curve $\boldsymbol{\phi}(\bold{x}_0,t)$; (b) the flow $\boldsymbol{\phi}(U,t)$ and for any $\bold{x} \in \mathbb{R}^n$ (\textit{i.e. $U=\mathbb{R}^n$}), the convention of representing the flow is $\boldsymbol{\phi}(\bold{x},t)$, which describes the time evolution of a class of initial points. This figure is motivated from \cite{guckenheimer2013nonlinear}.}\label{fig1}
\end{figure}

The Koopman operator is an infinite dimensional, linear, \textit{unitary operator} \cite{hoffmann1971linear,MR1129886} that acts on the Hilbert space of \textit{observables} \cite{powell1945dynamics} of the dynamical system.
Let a general smooth nonlinear dynamical system \cite{guckenheimer2013nonlinear} be defined as
\begin{equation}
    \centering
    \dot{\bold{x}}=\bold{f}(\bold{x}),
    \label{1}
\end{equation}
 where $\bold{x}=\bold{x}(t)\in \mathbb{R}^n$ is a vector-valued function of an independent variable (time) and the \textit{vector field}, $\bold{f} \colon U \rightarrow \mathbb{R}^n$ is a smooth function defined on an arbitrary open region $U \subseteq \mathbb{R}^n$; see Fig.\ref{fig1}(a). The vector field $\bold{f}$ generates a flow $\boldsymbol{\phi} \colon U \times I \rightarrow \mathbb{R}^n$. This implies that $\boldsymbol{\phi} (\bold{x}_0,t)$, assigns a point in the state space from an initial condition $\bold{x}_0 \in U$ and time $t \in I$, where $I$ is an open interval $(a,b) \in \mathbb{R}$ as presented in Fig.\ref{fig1}(a). The time evolution of the entire open region $U$ is represented as $\boldsymbol{\phi}(U,t)$, in Fig.\ref{fig1}(b). For the purpose of this article, the domain of definition of $\bold{f}$ is set as $\mathbb{R}^n$ ($U=\mathbb{R}^n$); and the time evolution of a class of initial points is conventionally represented as $\boldsymbol{\phi}(\bold{x},t)$.
%

Let us consider such functions, $g \colon \mathbb{R}^n \rightarrow \mathbb{C}$, that maps any element of the state space $\mathbb{R}^n$ to the complex plane $\mathbb{C}$. This choice of the co-domain of the measurement function $g$, is simply to keep it general, which makes it possible to analyze for the cases where the eigenvalues of the dynamical system are complex numbers. In the parlance of dynamical systems,  these functions are called \textit{observables} \cite{powell1945dynamics}.  
Let there be a vector space of observables $\mathcal{G}$ (Hilbert space). The Koopman operator $\mathcal{K}^t$ is defined as $\mathcal{K}^t \colon \mathcal{G} \rightarrow \mathcal{G}$, such that,
\begin{equation}
    \centering
    \mathcal{K}^tg(\bold{x})=g(\boldsymbol{\phi}(\bold{x},t)),
    \label{2}
\end{equation}
where, $g(\bold{x}) \in \mathcal{G}$ and $\boldsymbol{\phi}(\bold{x},t) \in \mathbb{R}^n$.
The fundamental property of the Koopman operator is linearity. Consider an observable $g_\gamma(\bold{x}) \in \mathcal{G}$, which is a linear combination of two observables $g_\alpha(\bold{x}) \ \text{and} \ g_\beta(\bold{x}) \in \mathcal{G}$. Mathematically, this implies 
\begin{equation}
    \centering
    g_\gamma(\bold{x})=a g_\alpha(\bold{x})+b g_\beta(\bold{x}).
    \label{4}
\end{equation}
Applying $\mathcal{K}^t$ on $g_\gamma(\bold{x})$ therefore leads to
\begin{equation}
    \centering
    \mathcal{K}^tg_\gamma(\bold{x})=g_\gamma(\boldsymbol{\phi}(\bold{x},t))=a g_\alpha(\boldsymbol{\phi}(\bold{x},t))+b g_\beta(\boldsymbol{\phi}(\bold{x},t))=a \mathcal{K}^tg_\alpha(\bold{x})+b \mathcal{K}^tg_\beta(\bold{x}).
    \label{3}
\end{equation}

\subsection{Eigenfunctions of the Koopman operator}
The Koopman operator is a linear transformation \cite{hoffmann1971linear,MR1129886}, $\mathcal{K}^t \colon \mathcal{G} \rightarrow \mathcal{G}$, thus have eigenvalues and corresponding eigenfunctions. The eigenfunctions of $\mathcal{K}^t$ are those observables $s(\bold{x}) \in \mathcal{G}$,  defined as
\begin{equation}
    \centering
    \mathcal{K}^ts(\bold{x})=s(\boldsymbol{\phi}(\bold{x},t))=s(\bold{x})\exp{(\lambda t)},
    \label{5}
\end{equation}
where, $\lambda \in \mathbb{C}$ (in general) is the corresponding eigenvalue of the eigenfunction $s(\bold{x})$.
The Koopman eigenfunctions have an interesting property, which enables constructing an eigenfunction from two different eigenfunctions. This can be mathematically stated as
\begin{equation}
    \centering
    \mathcal{K}^ts_{1,2}(\bold{x})=s_{1,2}(\boldsymbol{\phi}(\bold{x},t))=s_1(\boldsymbol{\phi}(\bold{x},t))s_2(\boldsymbol{\phi}(\bold{x},t))
    =s_1(\bold{x})s_2(\bold{x})\exp{((\lambda_1+\lambda_2) t)},
    \label{6}
\end{equation}
where $s_{1,2}(\bold{x})=s_{1}(\bold{x})s_{2}(\bold{x})$ is the constructed eigenfunction with eigenvalue $(\lambda_1+\lambda_2)$, from two different eigenfunctions $s_{1}(\bold{x})$ and $s_{2}(\bold{x})$ with corresponding eigenvalues $\lambda_1$ and $\lambda_2$.
In general, any linear combination of eigenvalues is an eigenvalue of the Koopman operator, and which corresponds to an eigenfunction of the Koopman operator. Thus, in general
\begin{equation}
    \centering
    s_{k_1,\ldots,k_n}(\bold{x})=s_1^{k_1}(\bold{x}) \cdots s_n^{k_n}(\bold{x})
    \label{7}
\end{equation}
and the corresponding eigenvalue is  $\lambda_{k_1,\ldots,k_n}=k_1\lambda_1+\cdots+k_n\lambda_n$; for $(k_1,\ldots,k_n)\in \mathbb{N}_0^n$.
Here $\mathbb{N}_0$ is the set of all \textit{non-negative integers}, defined as, $\mathbb{N}_0=\mathbb{N} \cup \{ 0 \}$, where $\mathbb{N}$ is the set of all natural numbers. These eigenvalues constitute the \textit{spectrum} of the Koopman operator \cite{kolmogorov1957elements}.

An additional important property embedded in the Koopman eigenfunctions is the invariance property of the \textit{zero level sets} of these eigenfunctions. Zero level sets are the set of points where a function yields zero values \cite{sethian1999level}. 
Let $s(\bold{x})$ be an eigenfunction and $\lambda$ be the corresponding eigenvalue. Then its zero level set $\{ \bold{x} \in \mathbb{R}^n | s(\bold{x})=0 \}$ is invariant for dynamics, as for any initial condition $s(\bold{x}_0)=0$, the corresponding trajectory will satisfy $s(\boldsymbol{\phi}(\bold{x}_0,t))=\exp{(\lambda t)}s(\bold{x}_0)=0$. Later in this paper, it will be shown that the invariance property of the Koopman eigenfunction is analogous to the invariant definition of NNMs and how this property plays a major role in finding NNMs.

\subsection{Spectral expansion of a linear system}
The motivation for applying the Koopman operator theory for nonlinear dynamical systems lies in 
its spectral properties. This 
enables applying principles of linear algebra to decouple the dynamics and represent the flow in terms of linear combinations of the modal dynamics.
Consider a linear system,
\begin{equation}
    \centering
    \dot{\bold{x}}=\bold{A}\bold{x}, \ \ \bold{x} \in \mathbb{R}^n
    \label{8}
\end{equation}
where $\bold{A}$ is a $n \times n$ diagonalizable matrix with distinct eigenvalues. As is well known, the corresponding flow can be represented in terms of the linear combination of the eigenvectors as
\begin{equation}
    \centering
    \boldsymbol{\phi}(\bold{x},t) = \sum_{k=1}^{n}\mathcal{P}_k( \bold{x} )\bold{v}_k\exp({\lambda_kt}),
    \label{9}
\end{equation}
where $\bold{v}_k$ is an eigenvector of $\bold{A}$ corresponding to the eigenvalue $\lambda_k$ and $\mathcal{P}_k(\bold{x})$ is the projection of $\bold{x}$ on to the eigenspace generated by $\bold{v}_k$, given by 
 

\begin{equation}
    \centering
    \mathcal{P}_k(\bold{x})=\langle \bold{x} | \bold{v}_k \rangle=\bold{w}_k^*\bold{x}.
    \label{16}
\end{equation}
Here $\langle \cdot | \cdot \rangle$ represents the inner product, $^*$ denotes the conjugate transpose and $\bold{w}_k$ are the left eigenvectors of $\bold{A}$. In other words, these are also the eigenvectors of $\bold{A}^*$.
As $\bold{A}$ has distinct eigenvalues, it can easily be proved (see Appendix \ref{secB})  that 
\begin{equation}
    \centering
    \bold{w}_k^*\bold{v}_j=\begin{cases}
    1, & k=j \\
    0, & k \neq j.
    \label{17}
    \end{cases}
\end{equation}
This orthonormality condition  leads to Eq.\eqref{16}. 
Applying the Koopman operator on $\mathcal{P}_k(\bold{x})$ therefore leads to
\begin{equation}
    \centering
    \mathcal{K}^t\mathcal{P}_k(\bold{x})=\mathcal{P}_k(\boldsymbol{\phi}(\bold{x},t))=\bold{w}_k^*\sum_{j=1}^{n}\mathcal{P}_j( \bold{x} )\bold{v}_j\exp({\lambda_jt})=\mathcal{P}_k(\bold{x})\exp{(\lambda_kt)}.
    \label{18}
\end{equation}
Eq.\eqref{18} proves that the projections $\mathcal{P}_k(\bold{x})$ are indeed the eigenfunctions of $\mathcal{K}^t$, corresponding to the eigenvalues $\lambda_k$, of the matrix $\bold{A}$. For the rest of this paper, the notation $s_k(\bold{x})$ (as, $\mathcal{P}_k(\bold{x}) \equiv s_k(\bold{x})$) is used for both the projections of the flow on the eigenspace of $\bold{A}$ and the eigenfunctions of the Koopman operator. Therefore 
\begin{equation}
    \centering
    \boldsymbol{\phi}(\bold{x},t) = \sum_{k=1}^{n}s_k( \bold{x} )\bold{v}_k\exp({\lambda_kt}),
    \label{9aa}
\end{equation}
is used to represent the flow.

\subsection{Vibrational modes, eigensubspaces and spectral expansion}
\begin{figure}[!h]
    \centering
    \includegraphics[width=0.6\textwidth]{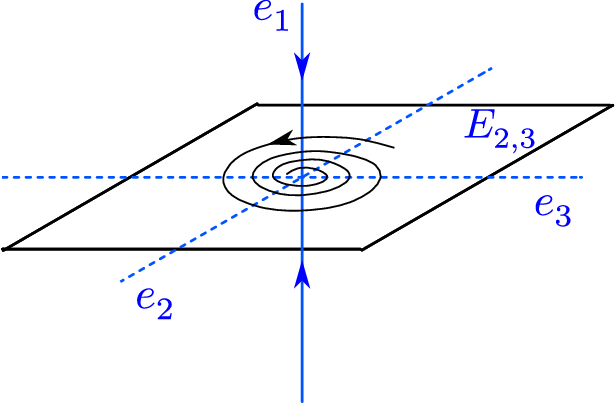}
    \caption{Schematic representation of the eigensubspaces of a $3$-dimensional linear dynamical system.}
    \label{LNM_3D}
\end{figure}

According to classical dynamical theory \cite{perko2013differential}, the LNMs of a system are defined as eigensubspaces. More precisely, LNMs corresponding to the ``\textit{real}'' eigenvalues of the system, are $1$-dimensional eigensubspaces, which essentially are the eigenvectors. Each LNM corresponding to the ``\textit{complex conjugate pair}'' of eigenvalues, is a $2$-dimensional eigensubspace spanned by the corresponding pair of complex conjugate eigenvectors. Fig.\ref{LNM_3D} shows a schematic representation of such a 2-dimensional eigensubspace for a $3$-dimensional 
linear dynamical system, typically represented as Eq.\eqref{8}, with $n=3$, \textit{i.e.} $\bold{x}(t) \in \mathbb{R}^3$, and which has a ``\textit{negative}'' (or ``\textit{positive}'') real eigenvalue $\lambda_1$ and a pair of complex conjugate eigenvalues $\lambda_2 \ \& \ \lambda_3$. The corresponding real eigenvector is $e_1$ and  $e_2 \ \& \ e_3$ represent the pair of complex conjugate eigenvectors.
The $2$-dimensional subspace spanned by $e_2 \ \& \ e_3$ can be represented as $E_{2,3}=\spn \{ e_2,e_3 \}$, as shown in Fig.\ref{LNM_3D}. Note the notion of the arrows in Fig.\ref{LNM_3D} implies stable dynamics. For unstable dynamics on any eigensubspace, the direction of the corresponding arrows will be reversed.

The general solution of the 3-dimensional dynamical system can be therefore represented mathematically as a linear combination of the dynamics on the $1$-dimensional LNM  $e_1$ and the dynamics on the $2$-dimensional LNM  $E_{2,3}$, as 
\begin{equation}
\label{9bb}
    \centering
    \bold{x}(t)=\overbrace{\mathcal{P}_1e_1\exp{(\lambda_1 t)}}^{\mathclap{\text{LNM-$1$: ``$1$-D eigensubspace'',} \ e_1}}+\underbrace{\mathcal{P}_2e_2\exp{(\lambda_2 t)}+\mathcal{P}_3e_3\exp{(\lambda_3 t)},}_{\text{LNM-$2$: ``$2$-D eigensubspace'',} \ E_{2,3}}
\end{equation}
where, $\mathcal{P}_j$ refers the projection of the initial state $\bold{x}(0)$ on the eigenvector $e_j, \  j= \{ 1,2,3 \}$. The dynamics on the $1$-dimensional LNM $e_1$ is trivial and linearly independent of the dynamics on the $2$-dimensional LNM $E_{2,3}$ and can  be determined separately. As the motion on $E_{2,3}$ is more interesting, most studies focus on $2$-dimensional subspaces (for LNMs) or $2$-dimensional manifolds (for NNMs). For the rest of this article, systems corresponding to the later are considered, where the dimension $n$ of the state space is assumed to be even ({\it i.e.}, $n=2m; m \in \mathbb{N}$).

Therefore, for the general $n$-dimensional ($n=2m; m \in \mathbb{N}$) system represented in Eq.\eqref{8}, the trajectory on any LNM plane  (defined as $\spn \{ \bold{v}_i,\bold{v}_j \}$) can be expressed as the linear combination of the dynamics along the corresponding eigenvectors only, and can be mathematically expressed as  $s_i(\bold{x})\bold{v}_i\exp{(\lambda_i t)}+s_j(\bold{x})\bold{v}_j\exp{(\lambda_j t)}$. Here, $\bold{v}_i$ and $\bold{v}_j$ are the eigenvectors corresponding to the pair of conjugate eigenvalues $\lambda_i$ and $\lambda_j$. For example, consider the pair of complex conjugate eigenvalues, $\lambda_1 \ {\rm and} \ \lambda_2$, the motion on the corresponding LNM plane can be represented as the linear combination of the dynamics along the eigenvectors $\bold{v}_1$ and $\bold{v}_2$ only. Thus, the right-hand side of Eq.\eqref{9aa} reduces to
\begin{equation}
    \centering
\sum_{k=1}^{n}s_k( \bold{x} )\bold{v}_k\exp({\lambda_kt})=s_1(\bold{x})\bold{v}_1\exp{(\lambda_1t)}+s_2(\bold{x})\bold{v}_2\exp{(\lambda_2t)},
\label{19}
\end{equation}
which corresponds to the zero-level set of all the other eigenfunctions \cite{cirillo2016spectral}. In other words,
\begin{equation}
    \centering
    s_k(\bold{x})=0, \ \forall k \in \mathbb{N} \setminus \{ 1,2 \}.
    \label{20}
\end{equation}
Thus, the plane ($2$- dimensional subspace) defining the LNM is determined by Eq.\eqref{20}. This property can be generalized to nonlinear systems and is equivalent to the invariant manifolds proposed by Shaw \textit{et. al.} \cite{shaw1991non,shaw1993normal}.
%
The NNM manifold has been  parameterized in \cite{cirillo2016spectral} through a nonlinear coordinate transformation, which involves mapping the state variables onto the vector space of observables $\mathcal{G}$. However, while the existence of an analytic manifold is asserted, no further proof of its uniqueness or criteria for existence is provided. Similarly, the literature lacks discussion on the proof of the uniqueness of mapping from the state space to $\mathcal{G}$. The subsequent sections present discussions on these significant gaps in the literature.
\section{Nonlinear coordinate transformation} \label{sec4}
Consider a nonlinear
dynamical system with continuously differentiable vector field 
defined in some open region $U$ of $\mathbb{R}^n$, whose flow field is given by
\begin{equation}
    \centering
    \dot{\bold{x}}=\bold{f}(\bold{x})= \bold{A}\bold{x}+\bold{\mathcal{N}}(\bold{x}),
    \label{21}
\end{equation}
where the origin $\bold{x}=\bold{0}$, is an equilibrium point (without loss of generality origin can be considered as a fixed point) contained in $U$, $\bold{A}=\frac{\partial \bold{f}}{\partial\bold{x}}\rvert_{\bold{x}=\bold{0}}$ 
is the Jacobian  of $\bold{f}$ evaluated at $\bold{x}=\bold{0}$ denoting the gradient of the vector field \cite{riley1999mathematical} 
and the function $\bold{\mathcal{N}}(\bold{x}) \sim \mathcal{O}(\| \bold{x} \|^2)$ accounts for the nonlinear variation of $\bold{f}$. The dynamical system represented by Eq.\eqref{21} induces a flow $\boldsymbol{\phi}(\bold{x},t) \colon U \times \mathbb{R} \rightarrow U$. As is well known, the local behaviour of the flow decides the stability of the equilibrium point   and is investigated by linearizing of the system in it its neighborhood. 


\subsection{Linearization in the neighborhood of the equilibrium }
Theorems on the local linearization of a nonlinear vector field, such as the \textit{Hartman-Grobman theorem} \cite{guckenheimer2013nonlinear,newhouse2017differentiable,lan2013linearization} the \textit{Hartman theorem} \cite{newhouse2017differentiable,lan2013linearization} and the \textit{Sternberg theorem} \cite{sternberg1957local,sternberg1957localreal,bonckaert1984linearization} suggest that for a dynamical system with hyperbolic fixed points, it is possible to attain a smooth change of co-ordinates in the neighborhood of the fixed point, such that the system behaves locally as a linear system.
%
For the sake of ease of exposition, the well known Hartman theorem is first stated here for convenience. Subsequently, we prove a  corollary, which supports the claim of Koopman eigenfunction to be a $C^1$-diffeomorphism map   from the neighborhood of $\bold{x}=\bold{0}$; see Appendix \ref{sec_def} for definitions.
\begin{theorem}[Hartman theorem]\label{thm2}
    Let $\bold{f} \in C^2(U)$. If all of the eigenvalues of the matrix $\bold{A}$ have negative (or positive) real part, then $\exists$ a $C^1$-diffeomorphism $\bold{d}=\bold{x}+\bold{\widetilde{d}}$ of a neighborhood $\mathcal{U} \subset U$ of $\bold{x}=\bold{0}$ onto an open set $\mathcal{V} \subset \mathbb{R}^n$ containing the origin, such that for each $\bold{x} \in \mathcal{U}$ there is an open interval $\mathcal{I}(\bold{x}) \subset \mathbb{R}$ containing zero and 
    $\forall \ \bold{x} \in \mathcal{U}$ and $t \in \mathcal{I}(\bold{x})$,
    \begin{equation}
        \centering
        \bold{d} \circ \boldsymbol{\phi}(\bold{x},t)=\exp{(\bold{A}t)}\bold{d}(\bold{x}).
     \label{24e}
    \end{equation}
    In addition,
    \begin{equation}
        \centering
        \lim_{\bold{x} \to \bold{0}} \frac{\| \bold{\widetilde{d}}(\bold{x}) \|}{\| \bold{x} \|}=\bold{0}, 
        \label{24f}
    \end{equation}
    where the norm of any vector $\boldsymbol{\alpha} \in \mathbb{R}^n$ is defined as, $\| \boldsymbol{\alpha} \| = \sqrt{\langle \boldsymbol{\alpha} | \boldsymbol{\alpha} \rangle}$.
\end{theorem}
Note that the Koopman operator $\mathcal{K}^t$ corresponds to Eq.\eqref{21}, acting on an observable $g \colon U \rightarrow \mathbb{C}$ is defined as $\mathcal{K}^t g(\bold{x})=g(\boldsymbol{\phi}(\bold{x},t))$. For the function (observable) $s(\bold{x})$ to be an eigenfunction of $\mathcal{K}^t$ associated with eigenvalue $\lambda$, it 
follows that
\begin{equation}
    \centering
    \mathcal{K}^t s(\bold{x})=s \circ \boldsymbol{\phi}(\bold{x},t)=\exp{(\lambda t)}s(\bold{x}). \label{24a}
\end{equation}

 \begin{corollary}\label{corl1}
 Using similar arguments, we could call the matrix $\bold{A}$ an eigenmatrix of $\mathcal{K}^t$ associated with eigenmapping $\bold{d}$, provided that {Eq.\eqref{24e}} holds. This diffeomorphism only holds locally {\it i.e.}, in the neighborhood of the origin.
 \end{corollary}
\proof[Corollary~{\upshape\ref{corl1}}]
 Let $\bold{A}$ have distinct eigenvalues, such that $\bold{A}=\boldsymbol{\Phi}^*\boldsymbol{\Lambda} \boldsymbol{\Phi}$, where   $\boldsymbol{\Lambda}$ is a diagonal matrix containing the eigenvalues and $\boldsymbol{\Phi}$ is the modal matrix whose columns are the eigenvectors.
 \footnote{Since the matrix $\bold{A}$ has distinct eigenvalues and here we are assuming it is diagonalizable, the modal matrix $\boldsymbol{\Phi}$, which consists of normalized eigenvectors, will have unitary property, \textit{i.e.}, $\boldsymbol{\Phi}^*=\boldsymbol{\Phi}^{-1}$, as the eigenvectors are orthonormal (see Appendix \ref{secB}).}. The linear transformation $\bold{z}=\boldsymbol{\Phi}^{*}\bold{x}$ leads to expressing the linearized equation in the ${\bold z}$-space as
 \begin{equation}
     \centering
     \dot{\bold{z}}=\boldsymbol{\Phi}^{*}\bold{A}\boldsymbol{\Phi}\bold{z} = \boldsymbol{\Lambda}\bold{z}.
     \label{24b}
 \end{equation}
  Pre-multiplying both sides of Eq. \eqref{24e} with $\boldsymbol{\Phi}^{*}$, leads to
  \begin{equation}
      \centering
      \boldsymbol{\Phi}^{*}\bold{d} \circ \boldsymbol{\phi}(\bold{x},t)=\boldsymbol{\Phi}^{*}\exp{(\bold{A}t)}\bold{d}(\bold{x}). \label{24c}
  \end{equation}
 Introducing $\bold{k}=\boldsymbol{\Phi}^{*}\bold{d}$  results in the simplified form
 \begin{equation}
     \centering
     \bold{k} \circ \boldsymbol{\phi}(\bold{x},t)=\boldsymbol{\Phi}^{*}\exp{(\bold{A}t)}\boldsymbol{\Phi}\bold{k}(\bold{x})=\exp{(\boldsymbol{\Lambda}t)}\bold{k}(\bold{x}),\label{24d}
 \end{equation}
where, each component of $\bold{k}$ is an eigenfunction of $\mathcal{K}^t$. \\
 \qed

We next state the Sternberg theorem \cite{sternberg1957local,sternberg1957localreal,bonckaert1984linearization} that ensures the existence of an analytic diffeomorphism (see Appendix \ref{sec_def}) in the neighborhood of $\bold{x}=\bold{0}$, and which provides the basis for nonlinear co-ordinate transformation of the dynamical system. 
 \begin{theorem}[Sternberg theorem]
\label{thm3}
    Let $\bold{f} \in C^{\infty}(U)$ vector field with $\bold{f}(\bold{0})=\bold{0}$. Suppose that,
    \begin{enumerate}
        \item each eigenvalue $\lambda_i$ of $\bold{A}$ satisfies $\Re (\lambda_i) < 0$ (hyperbolic fixed points);
        \item $\lambda_i \neq \sum_{j} m_j \lambda_j$ (non-resonant condition) for any non-negative integer $m_j \in \mathbb{N}_0$ such that, $\sum_{j} m_j > 1$;
    \end{enumerate}
    then there exists a unique local $C^{\omega}$-diffeomorphism (analytic diffeomorphism {\rm \cite{arnold2012geometrical}}) $\bold{z}$ in the neighborhood of the origin, such that the local change in co-ordinate can be represented as,
    \begin{equation}
    \centering
    \bold{z}=\bold{x}+\boldsymbol{\zeta}(\bold{x}),\ \ \ \ \boldsymbol{\zeta}(\bold{0})=\bold{0}, \ \ \ \ \left.\frac{\partial \boldsymbol{\zeta}}{\partial \bold{x}}\right\vert_{\bold{x}=\bold{0}}=\bold{0},
    \label{25}
\end{equation}
where the function $\boldsymbol{\zeta}(\cdot)$ follows,
\begin{equation}
    \centering
    \dot{\boldsymbol{\zeta}}(\bold{x})=\bold{A}\boldsymbol{\zeta}-\bold{\mathcal{N}}(\bold{x}).
    \label{26}
\end{equation}
\end{theorem}
This enables representing the  dynamics in $\bold{z}$ co-ordinate as
\begin{equation}
    \centering
    \dot{\bold{z}}=\bold{A}\bold{z}.
    \label{27}
\end{equation}
The Sternberg theorem establishes two essential criteria for the existence of analytic diffeomorphism, enabling local smooth coordinate changes in the neighbourhood of any fixed points of the system. First, it mandates that all the fixed points of the system are hyperbolic. 
This condition ensures that any non-hyperbolic fixed points (where the real part of the eigenvalues, $\Re(\lambda_i)=0$ for any $i \in \{ 1,\ldots,n \}$) result in the corresponding eigenspace in the fixed point's neighborhood being a central eigenspace ($E_c$). Consequently, the invariant manifold tangent to this eigenspace becomes a center manifold, as suggested by the center manifold theorem \cite{glendinning1994stability}, implying the manifold to be not unique. The second criterion states that the eigenvalues must be non-resonant, meaning that no eigenvalue can be expressed as the sum of multiples of other eigenvalues, \textit{i.e.}, $\lambda_i \neq \sum_{j} m_j \lambda_j$, for any non-negative integer $m_j \in \mathbb{N}_0$ such that, $\sum_{j} m_j > 1$. The presence of resonating eigenvalues results in the folding of NNM manifolds 
\cite{cirillo2016spectral}.

Lan and Mezi\'c \cite{lan2013linearization} extended this local smooth change of co-ordinates \textit{i.e.} diffeomorphism $\bold{z}(\bold{x})$ 
to the entire basin of attraction of the fixed point. This diffeomorphism is analytic only in the neighborhood of the origin and this smooth change of variable is only restricted for the system with non-resonant eigenvalues, (\textit{i.e.}, $\sum_j m_j \lambda_j \neq 0; \forall m_j \in \mathbb{N}_0, \sum_j m_j >1, j \in \{ 1,2,\ldots,n \} $). It has also been shown that the global diffeomorphism can only be $C^1$ function represented by $\bold{y}(\bold{x})$, which is analytic ($\bold{z}(\bold{x})$) in the neighborhood of the origin. Corollary \ref{corl1} ensures that the eigenfunction of the Koopman operator resembles such smooth change of variables (diffeomorphism) near the equilibrium point. In the same sense as Eq.\eqref{16} and Corollary \ref{corl1}, the Koopman eigenfunctions can be written as
\begin{equation}
    \centering
    s_k(\bold{x})=\langle \bold{y}(\bold{x}) | \bold{v}_k \rangle=\bold{w}_k^*\bold{y}(\bold{x}).
    \label{28}
\end{equation}
From Eq.\eqref{25}, it follows that
\begin{equation}
    \centering
    \left.\frac{\partial \bold{y}}{\partial \bold{x}}\right\vert_{\bold{x}=\bold{0}}=\bold{I},
    \label{29}
\end{equation}
which yields
\begin{equation}
    \centering
    \left.\frac{\partial s_k}{\partial \bold{x}}\right\vert_{\bold{x}=\bold{0}}=\bold{w}_k^*.
    \label{33}
\end{equation}
Here ${\bf I}$ is an identity matrix. This implies that the gradient of the eigenfunction at the origin is a left eigenvector of $\bold{A}$. The eigenfunctions $s_k(\bold{x})=\bold{w}_k^*\bold{y}(\bold{x})$ are related to the co-ordinate $\bold{z}$, as the diffeomorphism or the smooth change in variable $\bold{z}$ is the same as $\bold{y}(\bold{x})$, but only restricted to the neighborhood of origin. These eigenfunctions can also be written in the form
\begin{equation}
    \centering
    \boldsymbol{\xi}=\boldsymbol{\Phi}\bold{z}, \ \ {\rm where,} \ \ \boldsymbol{\Phi}= \begin{pmatrix}
        \bold{w}_1^* \\ \vdots \\ \bold{w}_n^*
    \end{pmatrix}.
    \label{33}
\end{equation}
 leading to the following form of the dynamical equation 
 \begin{equation}
     \label{33a}
     \centering
     \dot{\boldsymbol{\xi}}=\boldsymbol{\Phi}\bold{A}\boldsymbol{\Phi}^{-1}\boldsymbol{\xi}=\boldsymbol{\Lambda}\boldsymbol{\xi}.
 \end{equation}
 The benefit of this representation is that in the 
 $\boldsymbol{\xi}$ space, the dynamics is linear and hence the  principles of linear algebra can be used for analyzing the nonlinear system stated in Eq.\eqref{21}.

\section{Representation of state variable in Koopman spectrum}\label{sec5}

Representing $\bold{x}$ in terms of Koopman eigenfunction allows extending the Koopman operator theory for nonlinear systems as well. This is possible by mapping 
the state vector $\bold{x}$ 
to the infinite dimensional vector space of observables $\mathcal{G}$. Consider the vector space generated by the span of all the eigenfunctions of $\mathcal{K}^t$,  written as $\spn \{ s_{k_1,\ldots,k_n}(\bold{x})=\prod_{i=1}^n s^{k_i}_i(\bold{x}) \ \rvert \ k_i \in \mathbb{N}_0; \sum_{i=1}^n k_i > 0, i \in \{1,\ldots,n \} \}$, an infinite dimensional vector space over the base field $\mathbb{C}$. Let the vector space of observables
\begin{equation}
    \label{33b}
    \mathcal{G}=\spn \{ s_{k_1,\ldots,k_n}(\bold{x})=\prod_{i=1}^n s^{k_i}_i(\bold{x}) \ \rvert \ k_i \in \mathbb{N}_0; \sum_{i=1}^n k_i > 0, i \in \{1,\ldots,n \} \}.
\end{equation}
From the definition of spanning set \cite{hoffmann1971linear,MR1129886}, $\bold{x}$ can therefore be expressed as a linear combination of the eigenfunctions of the Koopman operator. This representation will be unique, if and only if the eigenfunctions form the basis of $\mathcal{G}$. From the definition of $\mathcal{G}$ in Eq.\eqref{33b}, the Koopman eigenfunctions forms a spanning set \cite{hoffmann1971linear} of $\mathcal{G}$. To prove that they are linearly independent, the following \textit{Proposition} is introduced along with the proof.
\vspace{0.3cm}
\begin{proposition}\label{pro1}
 Let $\bold{A}$ be a $n \times n$ diagonal matrix with distinct eigenvalues $\lambda_i; \ i=1,\ldots,n$, such that the real parts of the eigenvalues of $\bold{A}$ are all ``negative"  (stable dynamics) or all ``positive" (unstable dynamics \footnote{Exploring the case with all eigenvalues having ``positive" real parts confirms the \textit{Proposition}'s validity in such scenarios. This highlights the \textit{Proposition}'s robustness, showing applicability when all eigenvalues exhibit ``positive" real parts. Importantly, the \textit{Proposition} doesn't deem this case necessary for unstable dynamics; it simply asserts that a single eigenvalue with a ``positive" real part is sufficient for instability.}). Let $s_i(\bold{x}); \ i=1,\ldots,n$, be the eigenfunctions of the Koopman operator, corresponding to $\lambda_i$'s. Then all the Koopman eigenfunctions, $\prod_{i=1}^n s^{k_i}_i(\bold{x}); \ k_i \in \mathbb{N}_0, \ \sum_{i=1}^n k_i > 0$, are linearly independent.
 \end{proposition}
\proof[Proposition~{\upshape\ref{pro1}}]
The proposition states that the Koopman eigenfunctions are linearly independent. First, it will be proved that any arbitrary product combination ($s_\aleph(\bold{x})=\prod_{i=1}^n s^{k_i}_i(\bold{x}); \ \aleph>n \in \mathbb{N}_0, k_i \in \mathbb{N}_0, \ \sum_{i=1}^n k_i > 1$) of Koopman eigenfunctions are linearly independent. Subsequently,  using the same logic the linear independence of other $n$ number of eigenfunctions ($s_i(\bold{x}); \ i=1,\ldots,n$)  also follow. This statement will be proved by \textit{method of contradiction}. So the contradiction hypothesis will be, ``The eigenfunctions are linearly dependent''. 
\paragraph{}
Following from the contradiction hypothesis, it is safe to assume, that an arbitrary Koopman eigenfunction can be expressed as
\begin{equation}
    \label{9a}
    \centering
    c_1 s_1(\bold{x}) + c_2 s_2(\bold{x})+ \cdots + c_n s_n(\bold{x})+\cdots+ c_j s_j(\bold{x}) = c_\aleph s_\aleph(\bold{x}); \ \ c_i \in \mathbb{C},
\end{equation}
where $s_j(\bold{x})$ is a Koopman eigenfunction, obtained as the product of $s_i(\bold{x}); \ i=1,\ldots,n$, and $\lambda_j$ is the corresponding eigenvalue. By definition, all the eigenvalues of $\bold{A}$ are on the same half (positive or negative) of the complex plane. Therefore, without loss of generality  an ordering can be done such that,  
\begin{equation}
    \label{9b}
    \centering
    \Re(\lambda_1) > \Re(\lambda_2) > \cdots > \Re(\lambda_n).
\end{equation}
 Applying the Koopman operator on the right hand side of Eq.\eqref{9a} leads to
 \begin{align}
     \label{9c}
     \mathcal{K}^t (c_\aleph s_\aleph(\bold{x})) & =  \mathcal{K}^t (c_1 s_1(\bold{x}) + c_2 s_2(\bold{x})+ \cdots + c_j s_j(\bold{x})) \notag \\
     & = c_1 s_1(\bold{x}) \exp{(\lambda_1 t)} + c_2 s_2(\bold{x}) \exp{(\lambda_2 t)}+ \cdots + c_j s_j(\bold{x}) \exp{(\lambda_j t)} \notag \\
     & = c_\aleph s_\aleph(\bold{x}) \exp{(\lambda_\aleph t)} + \sum_{i=1}^j c_i s_i(\bold{x}) [\exp{(\lambda_i t)} -\exp{(\lambda_\aleph t)}].
 \end{align}
 From the definition of the Koopman operator eigenfunction, the left hand side of Eq.\eqref{9a} can be expressed as
 \begin{equation}
     \label{9d}
     \centering
     \mathcal{K}^t (c_\aleph s_\aleph(\bold{x})) = c_\aleph s_\aleph(\bold{x}) \exp{(\lambda_\aleph t)}.
 \end{equation}
 From Eq.\eqref{9c} and Eq.\eqref{9d}, it can be concluded that the \textit{contradiction hypothesis} (Eq.\eqref{9a}) is true iff the summation in Eq.\eqref{9c} is ``zero". As the eigenfunctions $s_\aleph(\bold{x})$ and $s_j(\bold{x})$ are obtained as the product of $s_i(\bold{x}); \ i=1,\ldots,n$, it follows that 
$s_\aleph(\bold{x})=\prod_{i=1}^n s^{k_i}_i(\bold{x}); \ \ k_i \in \mathbb{N}_0$ and $s_j(\bold{x})=\prod_{i=1}^n s^{k'_i}_i(\bold{x}); \ \ k'_i \in \mathbb{N}_0$, with the corresponding eigenvalues being
 \begin{equation}
     \label{9e}
     \centering
     \lambda_\aleph=\sum_{i=1}^n k_i \lambda_i \ \ {\rm and} \ \ \lambda_j=\sum_{i=1}^n k'_i \lambda_i.
 \end{equation}
As from Eq.\eqref{9e},
      $\text{either,}  \ \  \Re(\lambda_\aleph) > \Re(\lambda_j) > \Re(\lambda_1) > \Re(\lambda_2) > \cdots > \Re(\lambda_n) $ or 
     $\Re(\lambda_1) > \Re(\lambda_2) > \cdots > \Re(\lambda_n) > \Re(\lambda_j) > \Re(\lambda_\aleph)$,
it follows that
    \begin{equation}
        \label{9h}
        \centering     \sum_{i=1}^j[\exp{(\lambda_i t)} -\exp{(\lambda_\aleph t)}] \neq 0,
    \end{equation}
and therefore,
\begin{equation}
    \label{9i}
\centering
\mathcal{K}^t (c_\aleph s_\aleph(\bold{x})) \neq c_\aleph s_\aleph(\bold{x}) \exp{(\lambda_\aleph t)}. 
\end{equation}
    This is in direct contradiction to the definition of Koopman eigenfunction in Eq.\eqref{9d}. Thus, by the \textit{method of contradiction}, it is proved that all the product combinations of Koopman eigenfunctions ($\prod_{i=1}^n s^{k_i}_i(\bold{x}); \ k_i \in \mathbb{N}_0, \ \sum_{i=1}^n k_i > 1$) are linearly independent. The same logic follows for $j<n$ also, which completes the proof that all the Koopman eigenfunctions are linearly independent. \\
    \qed
 
 
\paragraph{}
 Proposition \ref{pro1} permits a unique representation of each component of $\bold{x}$, as a linear combination of eigenfunctions (basis of $\mathcal{G}$) of the Koopman operator, scaled with projections on the basis as 
\begin{equation}
    \centering
    x_i=\sum_{\substack{(k_1,\ldots,k_n) \in \mathbb{N}_0^n \\ k_1+ \cdots +k_n > 0}} (v_{k_1,\ldots,k_n})_is_1^{k_1}( \bold{x} ) \cdots s_n^{k_n}( \bold{x} ).
    \label{11}
\end{equation}
Representing the state variable in the above form 
enables finding an observable whose time evolution corresponds to the flow. The identity function $\bold{id}(\bold{x})=\bold{x}$, which is just the vector representation of Eq.\eqref{11}, allows such transformations as, $\mathcal{K}^t\bold{id}(\bold{x})  =\bold{id}(\boldsymbol{\phi}(\bold{x},t)) =\boldsymbol{\phi}(\bold{x},t)$. Using the same idea as Eq.\eqref{11} the identity function can be represented as 
\begin{equation}
    \centering
    \bold{id}(\bold{x})=\bold{x}=\sum_{\substack{(k_1,\ldots,k_n) \in \mathbb{N}_0^n \\ k_1+ \cdots +k_n > 0}} \bold{v}_{k_1,\ldots,k_n}s_1^{k_1}( \bold{x} ) \cdots s_n^{k_n}( \bold{x} ).
    \label{12}
\end{equation}
Eq.\eqref{12} is essentially a mapping $\mathcal{M} \colon \mathbb{R}^n \rightarrow \mathcal{G}$, where the $\bold{v}_{k_1,\ldots,k_n}$ are the projections on the basis of $\mathcal{G}$, \textit{i.e.} $\prod_{i=1}^n s^{k_i}_i(\bold{x}); \ k_i \in \mathbb{N}_0, \ \sum_{i=1}^n k_i > 0$. Here $\bold{v}_{k_1,\ldots,k_n}$ are elements of $\mathbb{R}^n$ and forms a spanning set. The inverse mapping from the range of $\mathbb{R}^n$, \textit{i.e.} $\mathrm{range}(\mathbb{R}^n)=G \subset \mathcal{G}$, is $\mathcal{M}^{-1} \colon G \rightarrow \mathbb{R}^n$, where $s_1^{k_1}( \bold{x} ) \cdots s_n^{k_n}( \bold{x} )$ are the projections on the spanning set $\bold{v}_{k_1,\ldots,k_n}$. It should be noted that only for linear systems $\bold{v}_{k_1,\ldots,k_n}$ forms the basis of $\mathbb{R}^n$, \textit{i.e.} only the linearly independent elements of $\bold{v}_{k_1,\ldots,k_n}$ remain nonzero. The time evolution of the identity function can be obtained by applying the Koopman operator,
leading to
\begin{equation}
   \begin{split}
        \centering
    \mathcal{K}^t\bold{id}(\bold{x}) & =\bold{id}(\boldsymbol{\phi}(\bold{x},t))=\boldsymbol{\phi}(\bold{x},t) \\ & =\sum_{\substack{(k_1,\ldots,k_n) \in \mathbb{N}_0^n \\ k_1+ \cdots +k_n > 0}} \bold{v}_{k_1,\ldots,k_n}s_1^{k_1}( \bold{x} ) \cdots s_n^{k_n}( \bold{x} )\exp{((k_1\lambda_1+\cdots +k_n\lambda_n)t)}.
    \label{13}
   \end{split}
\end{equation}
This is the flow generated by the vector field of the dynamical system in Eq.\eqref{1}.


\subsection{Relation between \textit{NNM}s and eigenfunctions of the Koopman operator} \label{sec5.1}

 Shaw \textit{et. al.} \cite{shaw1991non,shaw1993normal} defined \textit{NNM}s as $2$-dimensional invariant manifolds in state-space, which pass through a stable fixed point and is tangent to the linear normal mode of the linearized system at that point. In this section, it will be demonstrated, that the eigenfunctions of the Koopman operator satisfy all three criteria for \textit{NNM}s.
Eq.\eqref{19} describes the flow on the linear normal mode corresponding to the complex conjugate pair of eigenvalues $\lambda_1 \ {\rm and} \ \lambda_2$, and the mode constitutes the solution space of all the other eigenfunctions corresponding to the other eigenvalues; see Eq.\eqref{20}. This concept was generalized in \cite{cirillo2016spectral} for nonlinear systems by defining the NNMs as the surface ($2$-dimensional manifold embedded in $\mathbb{R}^n$) which corresponds to the solution space of  ($n-2$)  zero level sets of Koopman eigenfunctions,
\begin{equation}
    \centering
     s_k(\bold{x})=0, \ \ \forall k \in \mathbb{N} \setminus \{ 1,2 \}, \ \ \lambda_1=\overline{\lambda}_2.
     \label{34}
\end{equation}
\begin{enumerate}
    \item Invariance property: As on the manifold, the level sets $s_k(\bold{x})=0$ implies $s_k(\bold{x}_0)=0$, where $\bold{x}_0$ is some arbitrary initial condition on the manifold. If $\boldsymbol{\phi}(t,\bold{x}_0)$ describes the flow starting from this initial point 
    then from the definition of the eigenfunction of the Koopman operator (see Eq.\eqref{5}), one can write
    \begin{equation}
        \centering
        s_k(\boldsymbol{\phi}(t,\bold{x}_0))=\exp{(\lambda_k t)}s_k(\bold{x}_0)=0, \ \ \forall k \in \mathbb{N} \setminus \{ 1,2 \}, \ \  \forall t \in [0, \infty );
        \label{35}
    \end{equation}
    \textit{i.e.} the trajectory will always be on that manifold. This satisfies the invariance property.
    \item Contains the equilibrium point  (origin): From Eq.\eqref{25} and Eq.\eqref{28}, it follows that
    \begin{equation}
        \centering
        s_k(\bold{0})=\bold{w}_k^*\bold{y}(\bold{0})=\bold{w}_k^*\bold{0}=0.
        \label{36}
    \end{equation}
     As this implies that the origin `$\bold{0}$' satisfies the equation of any arbitrary level set $s_k(\bold{x})=0$, the intersection of these level sets also contains the origin.
    \item Tangent space of the manifold at the origin: According to the definition of tangent space (see Appendix \ref{secD}) of a manifold, the equation of tangent space at the origin is (refer to Eq.\eqref{appendixB7})
    \begin{equation}
        \centering
        \frac{\partial s_k}{\partial \bold{x}}(\bold{0})\bold{x}=0, \ 
        \ \forall k \in \mathbb{N} \setminus \{ 1,2 \}.
        \label{37}
    \end{equation}
    This implies that
    \begin{equation}
        \centering
        \bold{w}_k^*\bold{x}=0, \ 
        \ \forall k \in \mathbb{N} \setminus \{ 1,2 \},
        \label{38}
    \end{equation}
    which corresponds to the linear normal modes associated with $\lambda_1$ and $\lambda_2$.
\end{enumerate}

Therefore the solution space of the zero level sets $s_k(\bold{x})=0, \forall k \in \mathbb{N} \setminus \{ 1,2 \}$ satisfies all the conditions of the definition of NNM \cite{shaw1991non,shaw1993normal}.

\subsection{Parameterization of the invariant manifold} \label{NNM-para}

Based on the analysis in  Section \ref{sec5.1}, it is possible to use the methods of the Koopman operator to compute the NNMs.
Let $\xi_k=s_k(\bold{x})$ so that the dynamics in the $\boldsymbol{\xi}$ space can be expressed as (from Eq.\eqref{12})
\begin{equation}
    \centering
     \bold{id}(\bold{x})=\bold{x}=\sum_{\substack{(k_1,\ldots,k_n) \in \mathbb{N}_0^n \\ k_1+ \cdots +k_n > 0}} \bold{v}_{k_1,\ldots,k_n}\xi_1^{k_1}( \bold{x} ) \cdots \xi_n^{k_n}( \bold{x} ),
    \label{39}
\end{equation}
and the equation of zero-level sets of the eigenfunctions is given by
\begin{equation}
    \centering
     \xi_k=s_k(\bold{x})=0, \ \ \forall k \in \mathbb{N} \setminus \{ 1,2 \}, \ \ \lambda_1=\overline{\lambda}_2.
     \label{40}
\end{equation}
\begin{figure}[htbp]%
\centering
\includegraphics[width=0.9\textwidth]{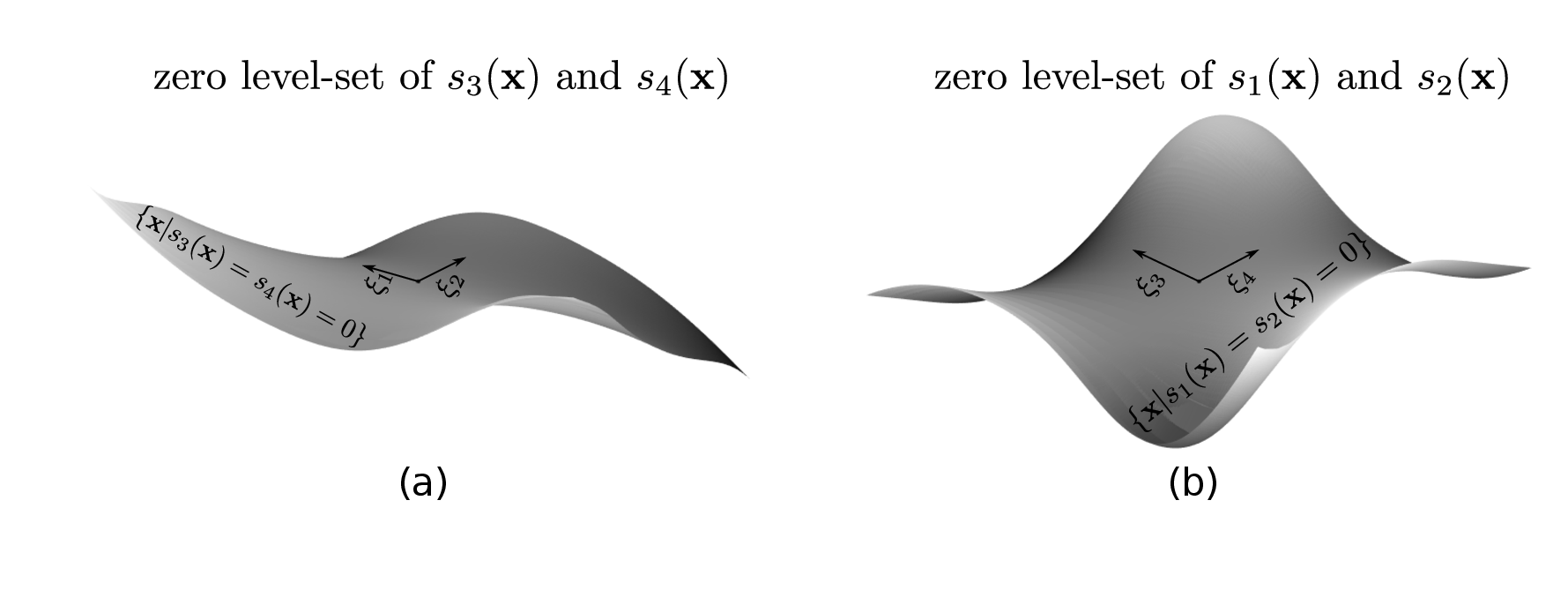}
\caption{Schematic diagram of two NNMs where the $2$-D manifold (a) corresponds to the zero level-set of $s_3(\bold{x})$ and $s_4(\bold{x})$, parameterized by $\xi_1 \ \& \ \xi_2$  and (b) corresponds to the zero level-set of $s_1(\bold{x})$ and $s_2(\bold{x})$, parameterized by $\xi_3 \ \& \ \xi_4$.}\label{fignum0}
\end{figure}
Eq. \eqref{39} represents the transformation of coordinates from $\bold{x}$ to $\boldsymbol{\xi}$ and Eq. \eqref{40} represents the NNM ($2$-D manifold) parameterized by $\boldsymbol{\xi}$. A schematic representation of such manifolds is shown in Fig.\ref{fignum0}. Fig.\ref{fignum0}(a) represents the zero level-set of $s_3(\bold{x})$ and $s_4(\bold{x})$, parameterized 
in terms of $\xi_1 \ \& \ \xi_2$, while  Fig.\ref{fignum0}(b) represents the zero level-set of $s_1(\bold{x})$ and $s_2(\bold{x})$, parameterized 
in terms of $\xi_3 \ \& \ \xi_4$.
So, the points on the manifold can be expressed as
\begin{equation}
    \centering
    \bold{x}=\sum_{\substack{(k_1,k_2) \in \mathbb{N}_0^2 \\ k_1+k_2 > 0}} \bold{v}_{k_1,k_2}\xi_1^{k_1}( \bold{x} ) \xi_2^{k_2}( \bold{x} ),
    \label{41}
\end{equation}
where the notation $\bold{v}_{k_1,k_2}=\bold{v}_{k_1,k_2,0,\ldots,0}$ is used. The time evolution of flow on the manifold is given by
\begin{equation}
    \label{42}
    \centering
    \bold{x}(t)=\boldsymbol{\phi}(\bold{x},t)=\sum_{\substack{(k_1,k_2) \in \mathbb{N}_0^2 \\ k_1+k_2 > 0}} \bold{v}_{k_1,k_2}\xi_1^{k_1}( \bold{x} ) \xi_2^{k_2}( \bold{x} )\exp{((k_1\lambda_1+k_2\lambda_2)t)}.
\end{equation}

\section{Systems with nonlinear internal resonance} \label{sec6}
\begin{figure}[htbp]%
\centering
\includegraphics[width=0.7\textwidth]{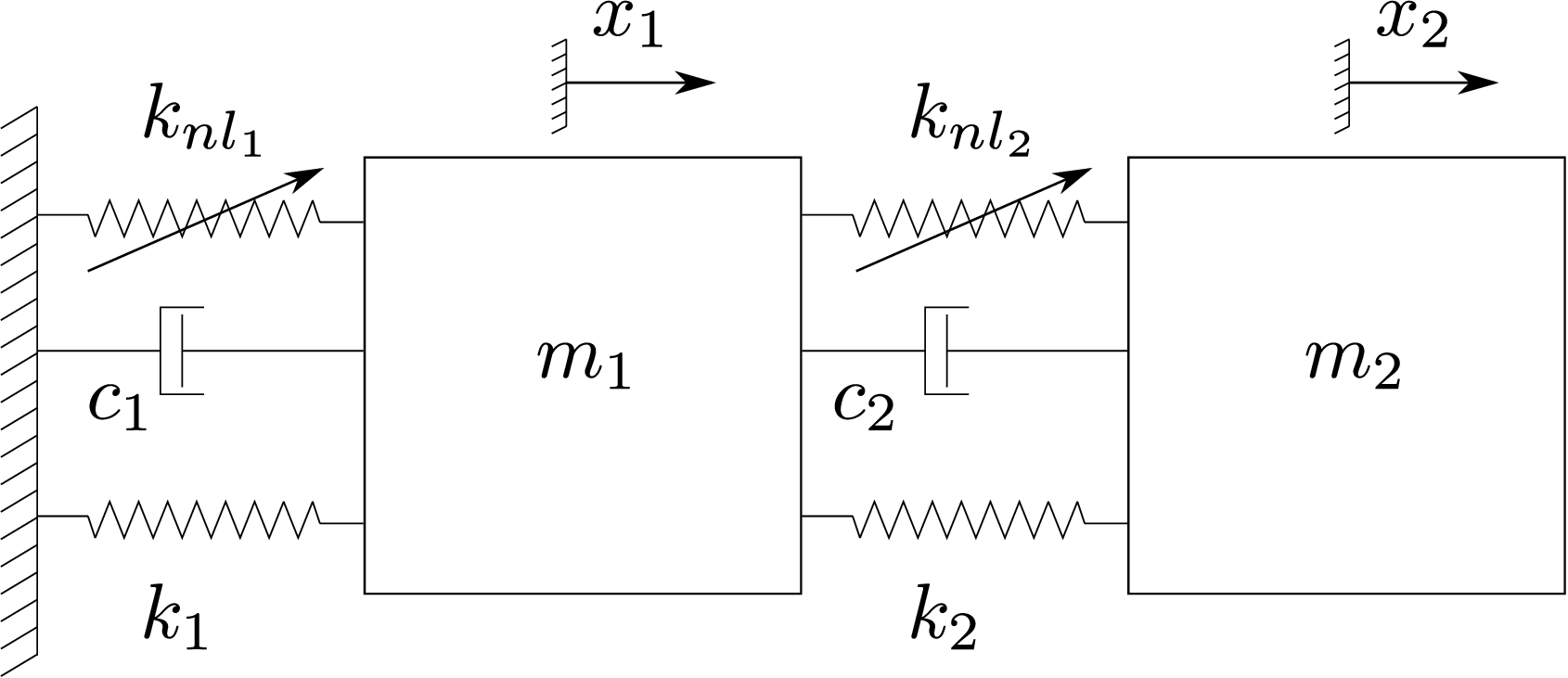}
\caption{Schematic diagram of two coupled Duffing oscillator.}\label{fig2}
\end{figure}

The Koopman operator theory appears to be a powerful tool for identifying NNMs with certain limitations, as Theorem \ref{thm3} states that, for resonating eigenvalues of the linearized system, the existence of analytic diffeomorphism in the neighborhood of the origin is lost. The analytic diffeomorphism ensures the uniqueness of the NNMs. Cirilo \textit{et. al.} \cite{cirillo2016spectral}
has explored  nonlinear systems with linear coupling. On the other hand, the existence of nonlinear coupling  sometimes lead to nonlinear internal resonance (even for non-resonating eigenvalues) due to
strong nonlinear interactions between the NNMs at higher energies. The motivation of this study is to explore such cases and observe their effects on the Koopman modes. We consider a $2$-dof system comprising of a Duffing oscillator coupled with another Duffing oscillator with the coupling stiffness having cubic nonlinearity; see Fig.\ref{fig2}.
Selecting such a system enables investigating the impact of internal resonance on the Koopman modes. The goal is to extend this theory to systems featuring essential nonlinearity as a limiting case. For this problem, the fixed points cease to be hyperbolic, leading to a failure of the Koopman operator framework. 

The equations of motion of this system, when normalized with respect to $m_1$, the mass of the first oscillator are 
\begin{align}
\label{44}
\Ddot{x}_1+\varepsilon\mu_1\dot{x}_1+\omega_0^2x_1+\varepsilon\alpha_1x_1^3+\varepsilon\mu_2(\dot{x}_1-\dot{x}_2)+\varepsilon \delta (x_1-x_2)+\varepsilon\alpha_2(x_1-x_2)^3=0 , \notag
\\
\Ddot{x}_2+\mu_2(\dot{x}_2-\dot{x}_1)+\delta (x_2-x_1)+\alpha_2(x_2-x_1)^3=0, 
\end{align}
where $\varepsilon=m_2/m_1$, $\mu_1=c_1/m_2$, $\omega_0^2=k_1/m_1$, $\alpha_1=k_{nl_1}/m_2$, $\mu_2=c_2/m_2$, $\delta=k_2/m_2$, and $\alpha_2=k_{nl_2}/m_2$, $m_1$, $m_2$ are the masses, $k_1$, $k_2$ are the linear part of the stiffness, $c_1$, $c_2$ are the damping coefficients, and $k_{nl_1}$, $k_{nl_2}$ are the cubic nonlinear stiffness of  oscillator $1$ and oscillator $2$ respectively. No restrictions are imposed on the 
mass ratio $\varepsilon$ and the linear part of the stiffness $\delta$; 
rather the effects of resonance is explored  by varying these parameters.
The damping coefficients for both oscillator have been considered to be equal, \textit{i.e.}, $\mu=\mu_1=\mu_2$.
 Eq.\eqref{44} can be compactly written in the matrix form as 
 \begin{equation}
    \centering
    \dot{\bold{x}}= \bold{A}
    \bold{x}
        +\begin{pmatrix}
        0 \\ 0 \\- \varepsilon(\alpha_1 x_1^3+\alpha_2(x_1-x_2)^3) \\ -\alpha_2(x_2-x_1)^3
    \end{pmatrix},
  \hspace{1.5cm}
    \label{45.5}
\end{equation}
 where, $\bold{x}=(x_1 \ x_2 \ \dot{x}_1 \ \dot{x}_2)^T$ and $\bold{A}=\begin{bmatrix}
    0 & 0 & 1 & 0 \\ 0 & 0 & 0 & 1 \\ -(\omega_0^2+\varepsilon\delta) & \varepsilon\delta & -2\varepsilon\mu & \varepsilon\mu \\ \delta & -\delta & \mu & -\mu
    \end{bmatrix}$.\\
    Differentiating Eq.\eqref{42} \textit{w.r.t.} time leads to 
    \begin{equation}
        \centering
        \dot{\bold{x}}=\frac{d \boldsymbol{\phi}(\bold{x},t)}{dt}=\sum_{\substack{(k_1,k_2) \in \mathbb{N}_0^2 \\ k_1+k_2 > 0}} \bold{v}_{k_1,k_2}\xi_1^{k_1} \xi_2^{k_2}(k_1\lambda_1+k_2\lambda_2)\exp{((k_1\lambda_1+k_2\lambda_2)t)}.
        \label{45.55}
    \end{equation}
    Expressing the vector $\bold{x}$ in the form of Eq.\eqref{41} and the vector $\dot{\bold{x}}$ in the form of Eq.\eqref{45.55}, the dynamical system  in 
 Eq.\eqref{45.5} can be represented in terms of the Koopman eigenfunctions and at $t=0$, is given by
\begin{equation}
        \centering
        \sum \bold{v}_{k_1,k_2}\xi_1^{k_1} \xi_2^{k_2}(k_1\lambda_1+k_2\lambda_2)=\bold{A}\sum \bold{v}_{k_1,k_2}\xi_1^{k_1} \xi_2^{k_2}+
        \begin{pmatrix}
            0 \\ 0 \\ \mathcal{NLT}_1 \\ \mathcal{NLT}_2
        \end{pmatrix}.
        \label{46}
    \end{equation}
Here, the summation symbol denotes $\sum \left( \cdot \right) \equiv \sum_{\substack{(k_1,k_2) \in \mathbb{N}_0^2 \\ k_1+k_2 > 0}} \left( \cdot \right)$, and the nonlinear terms are given by
    \begin{equation}
        \centering
        \mathcal{NLT}_1=-\varepsilon  \alpha_1 \left( \sum (\bold{v}_{k_1,k_2})_1\xi_1^{k_1} \xi_2^{k_2} \right)^3-\varepsilon\alpha_2\left( \sum \left( (\bold{v}_{k_1,k_2})_1- (\bold{v}_{k_1,k_2})_2 \right)\xi_1^{k_1} \xi_2^{k_2}\right)^3 ,
    \end{equation}
    \begin{equation}
        \centering
        \mathcal{NLT}_2=-\alpha_2 \left( \sum \left( (\bold{v}_{k_1,k_2})_2- (\bold{v}_{k_1,k_2})_1 \right)\xi_1^{k_1} \xi_2^{k_2} \right)^3,
    \end{equation}
    where, $(\bold{v}_{k_1,k_2})_i$ represents the $i^{\rm th}$ ($i \in \{1,\cdots,n\}$) element of the vector $\bold{v}_{k_1,k_2}$.
To determine the Koopman modes $\bold{v}_{k_1,k_2}$, Eq.\eqref{46} can be solved iteratively by comparing the order of $\xi_1^{k_1}\xi_2^{k_2}$, and leveraging the linear independence property of the Koopman eigenfunctions. 

\subsection{Solution Steps}
\label{sec6.1}

\begin{enumerate}
    \item[(a)] First order cases ($k_1+k_2=1$): 
    The equations for the first order Koopman modes are obtained by comparing the coefficients of $\xi_1^{k_1}\xi_2^{k_2}$ of both sides of Eq.\eqref{46}, such that $k_1+k_2=1$. This involves  the   cases  ($k_1=1, \ k_2=0$) or ($k_1=0, \ k_2=1$). 
    The corresponding linear equations are 
    \begin{equation}
    \label{47}
        \centering
        (\lambda_1 \bold{I}-\bold{A})\bold{v}_{10}=\bold{0}; \ \ (\lambda_2 \bold{I}-\bold{A})\bold{v}_{01}=\bold{0},
    \end{equation}
    where, $\bold{v}_{10}$ and $\bold{v}_{01}$ correspond to the first-order (linear) modes,
    and are obtained by finding out the \textit{null space} \cite{hoffmann1971linear,MR1129886} of $(\lambda_i \bold{I}-\bold{A}); \ i \in \{1,2\}$.
    
    \item[(b)] Second order cases ($k_1+k_2=2$): The equations for the second order cases are set up by comparing the coefficients of $\xi_1^{k_1}\xi_2^{k_2}$ on both sides of Eq.\eqref{46}, such that  $k_1+k_2=2$. 
    This results in three equations for the cases
    ($k_1=2, 
    \ k_2=0$) or ($k_1=1, \ k_2=1$) or ($k_1=0, \ k_2=2$). The corresponding equations are
    \begin{equation}
    \label{48}
        \centering
        (2\lambda_1 \bold{I}-\bold{A})\bold{v}_{20}=\bold{0}; \ \ ((\lambda_1+\lambda_2) \bold{I}-\bold{A})\bold{v}_{11}=\bold{0}; \ \ (2\lambda_2 \bold{I}-\bold{A})\bold{v}_{02}=\bold{0}.
    \end{equation}
    As the system considered in Eq.\eqref{44} do not have any 
    quadratic terms, 
    the second order modes are  zeros, \textit{i.e.} $\bold{v}_{20}=\bold{v}_{11}=\bold{v}_{02}=\bold{0}$.
    
    \item[(c)] Third order cases ($k_1+k_2=3$): The equations for the third order Koopman modes are similarly constructed 
    by comparing the coefficients of $\xi_1^{k_1}\xi_2^{k_2}$ from both sides of Eq.\eqref{46}, such that $k_1+k_2=3$. 
    This leads to the following four cases,  ($k_1=3, \ k_2=0$),  ($k_1=2, \ k_2=1$),  ($k_1=1, \ k_2=2$) and ($k_1=0, \ k_2=3$). The corresponding  equations are
    \begin{align}
        \label{49}
        (3\lambda_1 \bold{I}-\bold{A})\bold{v}_{30}  =[0,  0,  &-\varepsilon\alpha_1((\bold{v}_{10})_1^3-\varepsilon\alpha_2((\bold{v}_{10})_1-(\bold{v}_{10})_2)^3, \notag \\ 
        &-\alpha_2((\bold{v}_{10})_2-(\bold{v}_{10})_1)^3]^T; \notag \\
        ((2\lambda_1+\lambda_2 )\bold{I}-\bold{A})\bold{v}_{21}&=\bold{0}; \ \ ((\lambda_1+2\lambda_2 )\bold{I}-\bold{A})\bold{v}_{12}=\bold{0}; \notag \\
        (3\lambda_2 \bold{I}-\bold{A})\bold{v}_{03}  =[0,  0,  &-\varepsilon\alpha_1((\bold{v}_{01})_1^3-\varepsilon\alpha_2((\bold{v}_{01})_1-(\bold{v}_{01})_2)^3, \notag \\ 
        &-\alpha_2((\bold{v}_{01})_2-(\bold{v}_{01})_1)^3]^T.
    \end{align}
    \end{enumerate}
    It follows from Eqs.\eqref{47}-\eqref{49}  
    that the only nonzero modes are $\{\bold{v}_{10}$, $\bold{v}_{01}$, $\bold{v}_{30}$, $\bold{v}_{03}$, $\bold{v}_{90}$, $\bold{v}_{09}$, $\ldots \}$, or in general, $\{ \bold{v}_{r0},\bold{v}_{0r} \}; \ {\rm where,} \ r \in \{3^p \ | \ p \in \mathbb{N}_0 \}$. The NNM corresponding to the zero level-set of the eigenfunctions $s_3(\bold{x})$ and $s_4(\bold{x})$, shown in the Fig.\ref{fignum0}, can be constructed, using $\{\bold{v}_{r0}, \bold{v}_{0r} \}$. With the same procedure, the other NNM (zero level-set of the eigenfunctions $s_1(\bold{x})$ and $s_2(\bold{x})$) can be constructed using $\{ \bold{w}_{r0},\bold{w}_{0r} \}; \ r \in \{3^p \ | \ p \in \mathbb{N}_0 \}$. Here, $\bold{w}_{k_3k_4}$ are obtained from the equation
    \begin{equation}
        \centering
        \sum  \bold{w}_{k_3,k_4}\xi_3^{k_3} \xi_4^{k_4}(k_3\lambda_3+k_3\lambda_3)=\bold{A}\sum  \bold{w}_{k_3,k_4}\xi_3^{k_3} \xi_4^{k_4}+
        \begin{pmatrix}
            0 \\ 0 \\ \mathcal{NLT}_3 \\ \mathcal{NLT}_4
        \end{pmatrix},
        \label{50}
    \end{equation}
where the nonlinear terms are
\begin{equation}
        \centering
        \mathcal{NLT}_3=-\varepsilon  \alpha_1 \left( \sum  (\bold{w}_{k_3,k_4})_1\xi_3^{k_3} \xi_4^{k_4} \right)^3-\varepsilon\alpha_2\left( \sum \left( (\bold{w}_{k_3,k_4})_1- (\bold{w}_{k_3,k_4})_2 \right)\xi_3^{k_3} \xi_4^{k_4}\right)^3 ,
    \end{equation}
    \begin{equation}
        \centering
        \mathcal{NLT}_4=-\alpha_2 \left( \sum \left( (\bold{w}_{k_3,k_4})_2- (\bold{w}_{k_3,k_4})_1 \right)\xi_3^{k_3} \xi_4^{k_4} \right)^3.
    \end{equation}
    The Koopman modes $\{ \bold{v}_{r0},\bold{v}_{0r}, \bold{w}_{r0},\bold{w}_{0r} \}$ form the spanning set of the state space, \textit{i.e.}, $\spn \{\bold{v}_{r0},\bold{v}_{0r}, \bold{w}_{r0},\bold{w}_{0r} \} = \mathbb{R}^4$, where, $ r \in \{3^p \ | \ p \in \mathbb{N}_0 \}$. Note that the state space is $4$-dimensional for this example. 

\section{Numerical results} \label{sec7}
A numerical  parametric study is next carried out on the system described in Eq.\eqref{44}, 
with the following objectives:
\begin{enumerate}
    \item[(a)] To investigate and comprehend the characteristics of Koopman modes in the vicinity of nonlinear internal resonance regime.
    \item[(b)] To study the effects of the damping on the accuracy of the Koopman mode analysis.
    \item[(c)] To investigate the existence and the behaviour of Koopman modes, when the substructure is arbitrarily close to being an \textit{essentially nonlinear oscillator}. In other words,  this happens when  one of the eigenvalues of the system is arbitrarily close to zero.  Note that for an essentially nonlinear oscillator, one of the eigenvalues is zero and the methodology discussed in this paper is no longer valid.
\end{enumerate}
 The investigation proceeds by truncating terms beyond the order of $3^6$, \textit{i.e.} the Koopman modes being considered are $\{ \bold{v}_{r0},\bold{v}_{0r}, \bold{w}_{r0},\bold{w}_{0r} \}$; where $ r \in \{3^p \ | \ p=0,1,2,3,$ $4,5,6 \}$. This is feasible only due to the sparsity characteristics discussed in Section\ref{sec6.1}. The sparsity nature ensures that only $7$ terms need to be computed for this approximation. The numerical values for the parameters considered are
 $\omega_0=1$,  $\varepsilon \in \{ 0.2,0.3,\ldots,0.9,1 \}$, $\delta \in \{ 0.1,0.2,\ldots,0.9,1 \}$, $\mu \in \{ 0.1,0.2,\ldots,2,2.1 \}$, $\alpha \in \{ 2,3,\ldots,9,10 \}$ and $\beta \in \{ 2,3,\ldots,9,10 \}$. This leads to 
 $ \sim 2 \times 10^5$ parameter combinations. The statistical observation of the effects of internal resonance and damping ratio on the accuracy of the Koopman modes framework has been conducted across the mentioned parameter regime by defining a continuous quantitative measure for internal resonance. The numerical integrations have been cariied out using adaptive Runge-Kutta algorithm (ODE45) in \textsc{Matlab} \cite{MATLAB}.
 
 \subsection{Quantification of internal resonance} \label{sec7.1}
 
 Internal resonance is typically discussed in the literature \cite{DAS2023104285,vakakis2008nonlinear,kerschen2014modal,lacarbonara2003resonant}  in  a binary manner - either resonance capture or escape from resonance. However, empirical quantitative measures are occasionally deemed necessary for a more in-depth analysis. In this article, the instantaneous frequency ($\Omega(x_i)$) of the oscillators have been computed using the Hilbert transform \cite{luo2009hilbert}, and a metric is defined for continuous quantification of internal resonance between two substructures, based on $L^2$ norm ($\lVert \cdot \rVert _2$) \textit{i.e.}, the \textit{Euclidean distance} between two points on $\mathbb{R}^n$. The distance function 
 \begin{equation}
 \mathcal{D}(\bold{a},\bold{b})=\frac{1}{N} \sum_{i=1}^{N} \lVert a_i-b_i \rVert_2 \ ,
 \end{equation}
 measures the average distance of all the points on the $(\bold{a},\bold{b})$ plane from the $\bold{a}=\bold{b}$ line.
 \begin{figure}[!h]
\centering
\begin{minipage}[b]{0.45\textwidth}
\subcaptionbox{ \label{fignum2a}}{\includegraphics[width=0.8\textwidth]{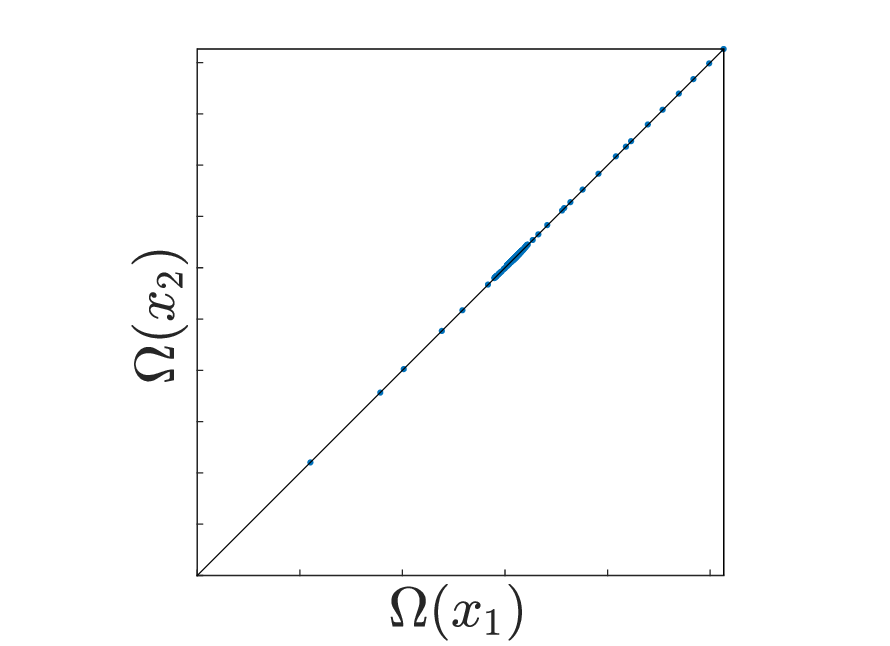}}    \vfill
\subcaptionbox{ \label{fignum2b}}{\includegraphics[width=0.8\textwidth]{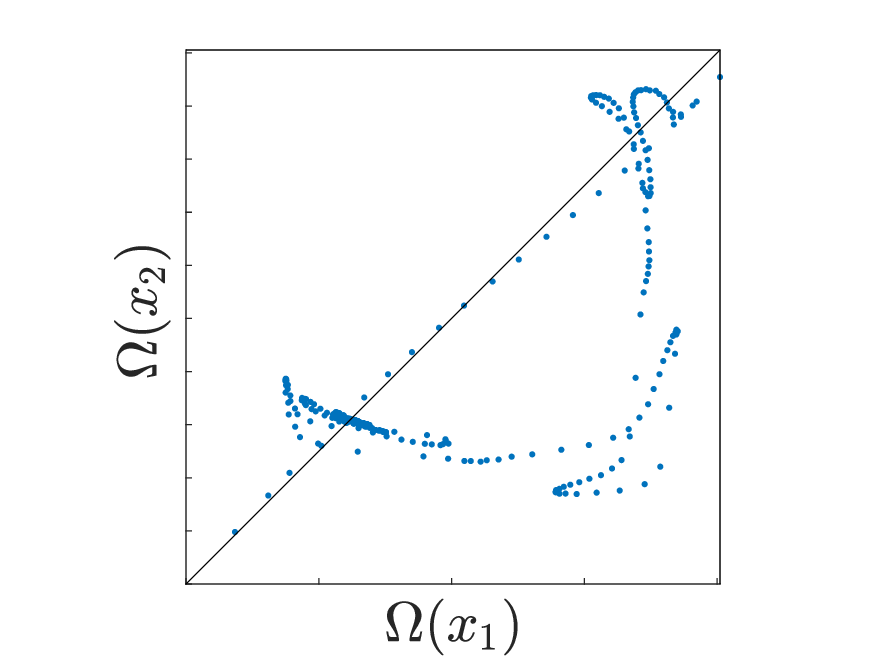}} \vfill
\subcaptionbox{ \label{fignum2c}}{\includegraphics[width=0.8\textwidth]{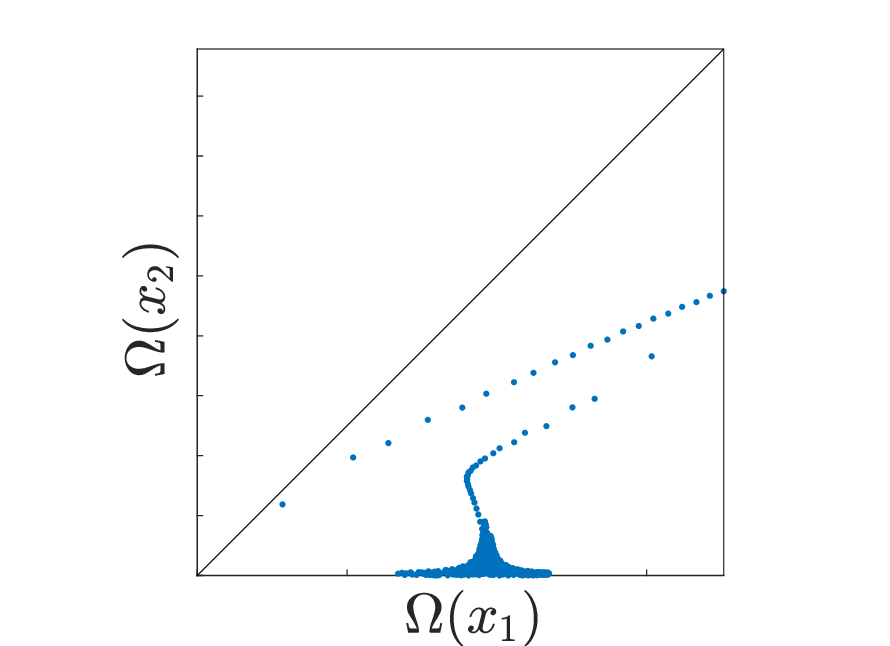}}
\end{minipage}
 \begin{minipage}[b]{0.45\textwidth}
\subcaptionbox{ \label{fignum2d}}{\includegraphics[width=0.8\textwidth]{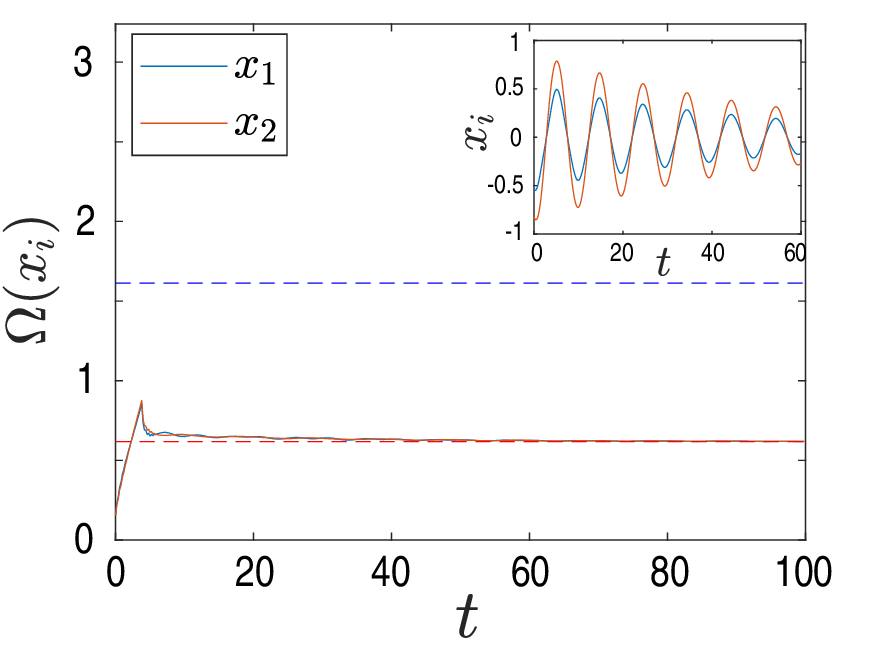}} \vfill
\subcaptionbox{ \label{fignum2e}}{\includegraphics[width=0.8\textwidth]{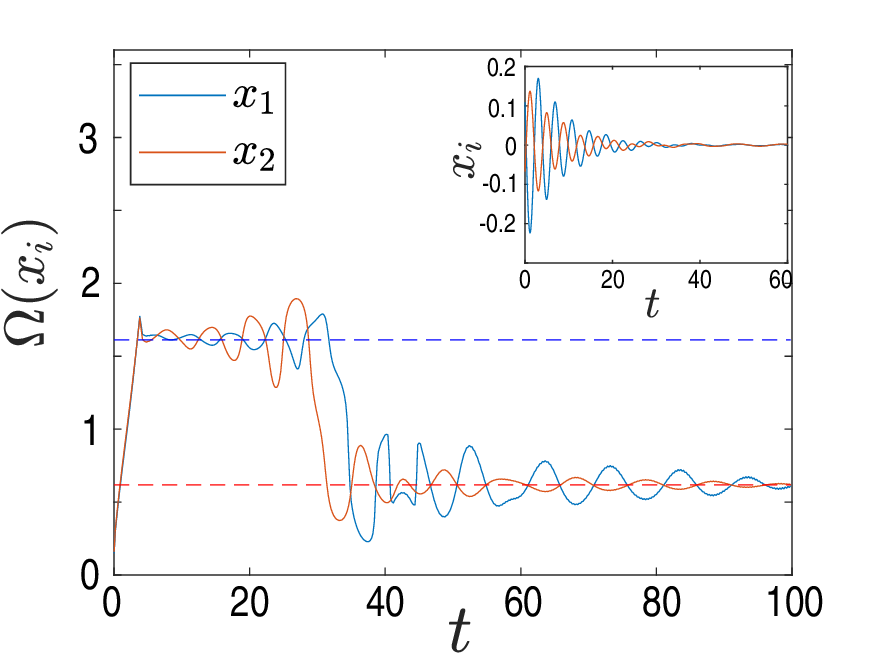}} \vfill
\subcaptionbox{ \label{fignum2f}}{\includegraphics[width=0.8\textwidth]{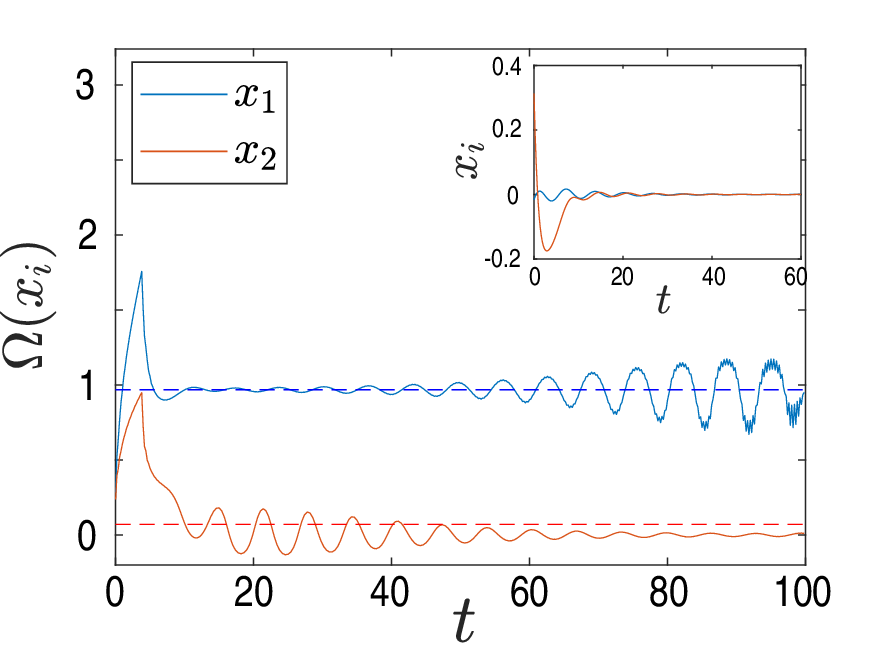}}
 \end{minipage}
\caption{Empirical representation for classifying three distinct cases of nonlinear internal resonance: (a) complete internal resonance, (b) near resonance and (c) no resonance, with corresponding instantaneous frequency plots: (d), (e) and (f).}
\label{fignum2}
\end{figure}
 \paragraph{}
 When both substructures are in a complete resonating condition, the instantaneous frequencies of each oscillator become exactly equal, \textit{i.e.}, all the points scattered on the graph $(\Omega(x_1),\Omega(x_2))$ shown in Fig.\ref{fignum2}(a), lie precisely on the $\Omega(x_1)=\Omega(x_2)$ line, indicating the point wise distance equal to zero, thus the mean distance of $\mathcal{D}( \Omega(x_1)-\Omega(x_2))=0$. Fig.\ref{fignum2}(d) illustrates the time evolution of the instantaneous frequencies of both oscillator ($\Omega(x_i); i \in \{ 1,2 \}$), along with the time histories of $x_i(t); i \in \{ 1,2 \}$ shown in the inset. This figure confirms the ``complete resonance capture'' between both the oscillators, since the instantaneous frequencies of both oscillators ($\Omega(x_i); i \in \{ 1,2 \}$) are exactly equal for all time. In Fig.\ref{fignum2}(b) most of the scattered points are in close proximity to the $\Omega(x_1)=\Omega(x_2)$ line, and the mean distance from the $\Omega(x_1)=\Omega(x_2)$ line will be non-zero. From Fig.\ref{fignum2}(e) it is  observed that the instantaneous frequencies of both oscillator, $\Omega(x_i); i \in \{ 1,2 \}$, are very close for all time, suggesting that the system exhibits ``near resonance behavior''. On the contrary, in Fig.\ref{fignum2}(c), all the scattered points are relatively distant from the $\Omega(x_1)=\Omega(x_2)$ line, resulting in a higher mean distance compared to the case  of ``near resonance''. Fig.\ref{fignum2}(f) confirms that the instantaneous frequencies of both the oscillators are not equal at any instant, indicating ``no resonance'' between the substructures. Therefore, this metric provides an intuitive assessment of the  proximity of the system to internal resonance. 
\subsection{Error estimation of the analytical results}
 The effects of damping and internal resonance on Koopman operator-based NNM  have been quantified through the  \textit{root mean square} measure, defined as
 \begin{equation}
 \sigma(x_i)=\frac{1}{N}\sqrt{\sum_{j=1}^{N}((x_i)_j-(\hat{x}_i)_j)^2} \ ,
 \end{equation}
where, $\sigma(x_i)$ 
is a measure for the error between the analytical approximation $x_i$, obtained from Eq.\eqref{42} 
and the numerical results $\hat{x}_i$, 
obtained by numerically integrating the ODEs represented in Eq.\eqref{44},  
and, $N$  is the number of time steps in the numerical integration. This measure serves as the foundation for exploring the system across a broad parametric range, encompassing approximately $ 2 \times 10^5$ cases, facilitating the observation of statistical trends. 

\begin{figure}[htbp]
\centering
\subcaptionbox{ \label{fignum3a}}{\includegraphics[width=0.45\textwidth]{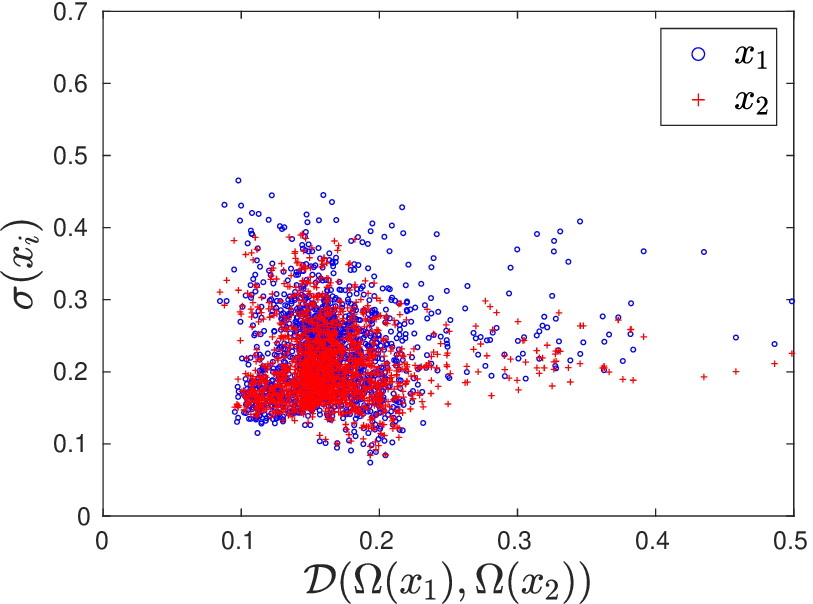}} \hfill
\subcaptionbox{ \label{fignum3b}}{\includegraphics[width=0.45\textwidth]{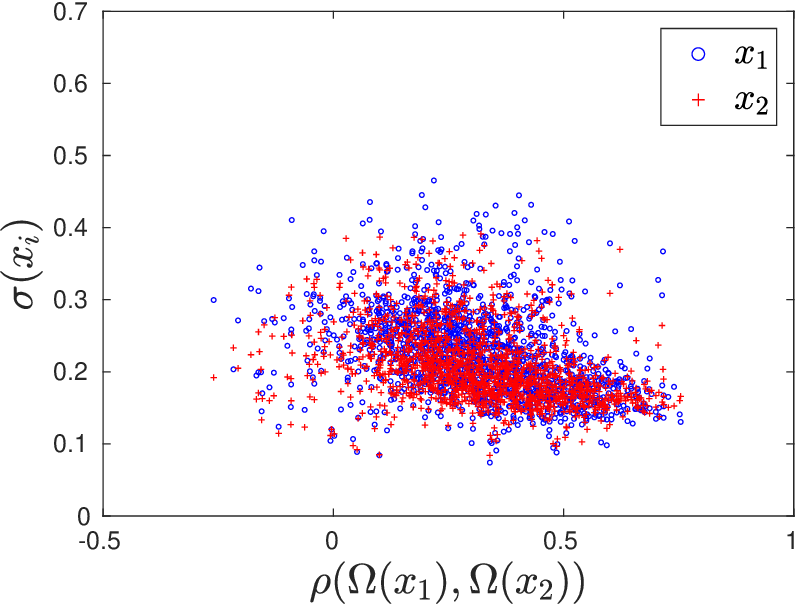}} 
\caption{Accuracy of analytical results at different resonance regime: (a) resonance metric ($\mathcal{D}(\Omega(x_1),\Omega(x_2))$) vs. $\sigma(x_i)$ and (b) correlation between instantaneous frequency $\rho(\Omega(x_1),\Omega(x_2))$ vs. $\sigma(x_i)$. The damping coefficient is taken to be $c=c_1=c_2=\varepsilon \cdot \mu=0.02$.}
\label{fignum3}
\end{figure}

Fig.\ref{fignum3}(a) presents a scatter plot illustrating the statistical trend of the root mean square error $\sigma(x_i)$ associated with the analytical approximation within the Koopman operator framework, plotted against the resonance metric $\mathcal{D}(\Omega(x_1),\Omega(x_2))$. Each point on the scatter corresponds to the error $\sigma(x_i)$ relative to the distance $\mathcal{D}(\Omega(x_1),\Omega(x_2))$ from the so called ``complete internal resonance'' state. 
The errors for the approximation of $x_1$ are represented as blue circles (`$\circ$'), while those for $x_2$ are specified with red cross (`$+$'). Notably, the scatter indicates that as the system approaches resonance, characterized by a smaller distance metric $\mathcal{D}(\Omega(x_1),\Omega(x_2))$, the root mean square error $\sigma(x_i)$ tends to be lower. 
Fig.\ref{fignum3}(b) shows the statistical trend for the variation of $\sigma(x_i)$ as a function of
correlation between instantaneous frequencies $\rho(\Omega(x_1),\Omega(x_2))$, defined as
\begin{equation}
\rho(\Omega(x_1),\Omega(x_2)) = {\frac{\sum_{j} (\Omega(x_1)_j-\overline{\Omega(x_1)})(\Omega(x_2)_j-\overline{\Omega(x_2)})}{\sqrt{\sum_{j} (\Omega(x_1)_j-\overline{\Omega(x_1)})^2\sum_{j} (\Omega(x_2)_j-\overline{\Omega(x_2)})^2}},}
\label{correl}
\end{equation}
{ where the mean instantaneous frequency, $\overline{\Omega(x_i)}=\frac{1}{N}\sum_j \Omega(x_i)$, $i \in \{1,2\}$, and the summation index $j=1,\ldots,N$; $N$ being the number of time steps in the numerical calculation.}
It can be clearly seen that as the correlation $\rho(\Omega(x_1),\Omega(x_2))$ increases, $\sigma(x_i)$ decreases, providing a more direct comparison. 

\begin{figure}[htbp]
    \centering
    \includegraphics[width=0.45\textwidth]{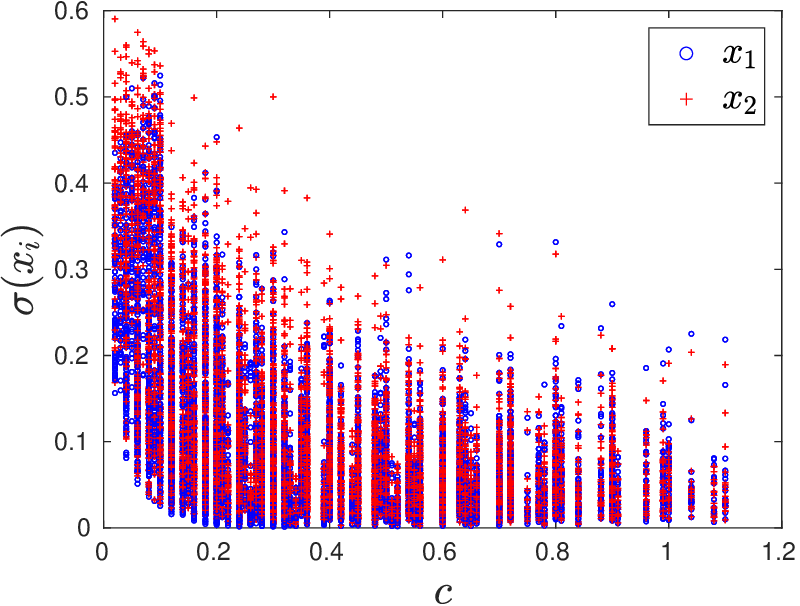}
    \caption{Error estimation at different damping values ($c$).}
    \label{fignum4}
\end{figure}

Figure\ref{fignum4} shows the impact of overall damping ($c=\varepsilon \cdot \mu$) on Koopman modes. It  is observed that at lower damping levels the error is notably high, but with increase in damping, the error decreases significantly. 
 This observation is intuitive, since damping contributes to the real part of the system's eigenvalues. A smaller damping parameter brings the eigenvalues closer to the non-hyperbolic case, thereby reducing the accuracy of the proposed framework. Additionally, at higher energy levels farther from the equilibrium point, the curvature of the manifold tends to be higher, because of the presence of higher order nonlinear terms in the parameterization of NNMs. With high damping, the energy of the system dissipates rapidly, and the system dynamics is mostly confined 
in the vicinity of the fixed point, where the NNM are comparatively flat. 

\begin{figure}[htbp]%
\centering
\includegraphics[width=0.9\textwidth]{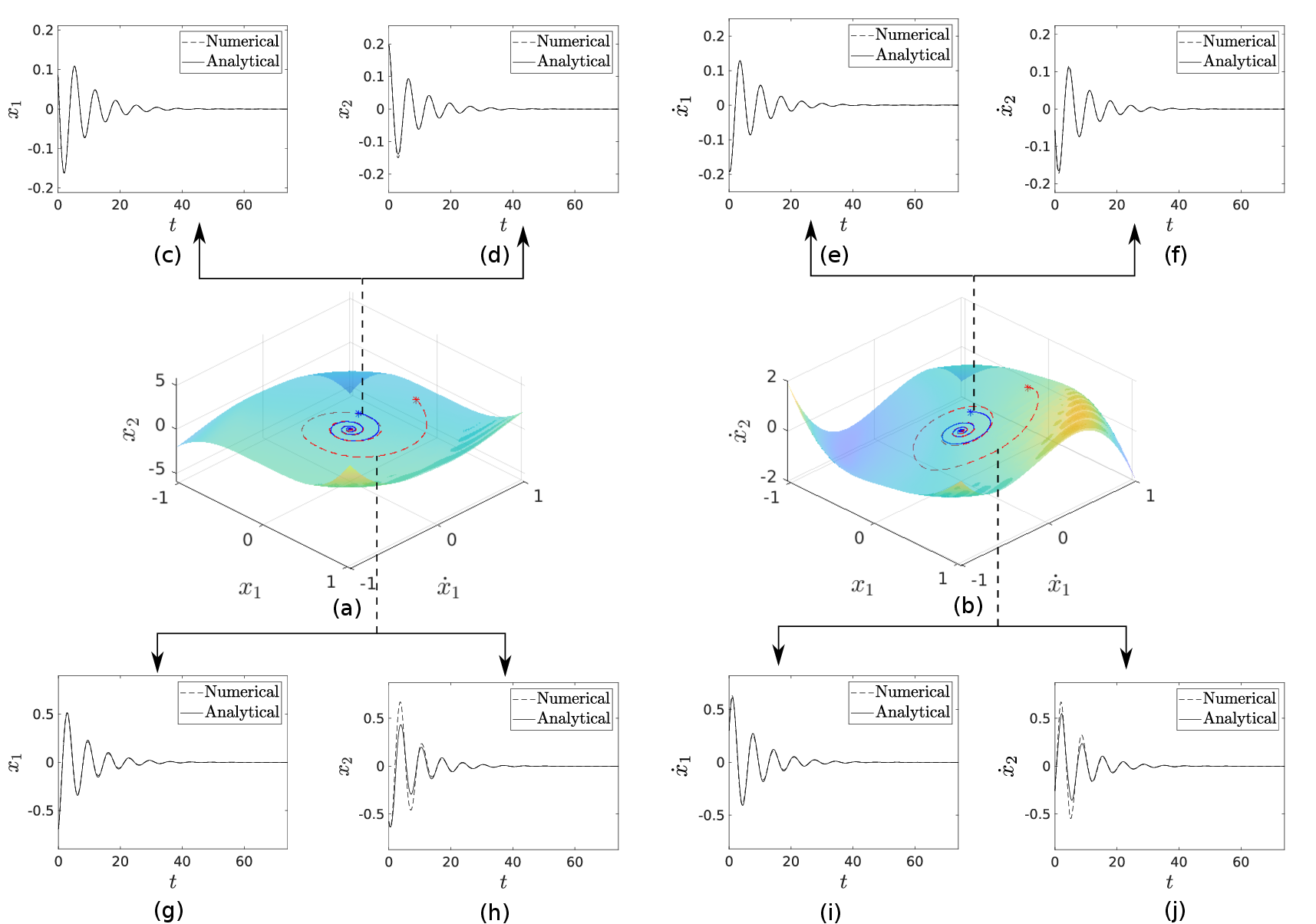}
\caption{NNM constructed with the Koopman modes: The projections of the invariant manifold on (a) the $x_1\dot{x}_1x_2$ space and (b) the $x_1\dot{x}_1\dot{x}_2$ space, are shown with the trajectories from numerical data. The time histories of all state variables: (c)-(f) correspond to the lower energy trajectories and (g)-(j) correspond to the higher energy trajectories.}\label{fignum5}
\end{figure}

\begin{figure}[htbp]%
\centering
\includegraphics[width=0.9\textwidth]{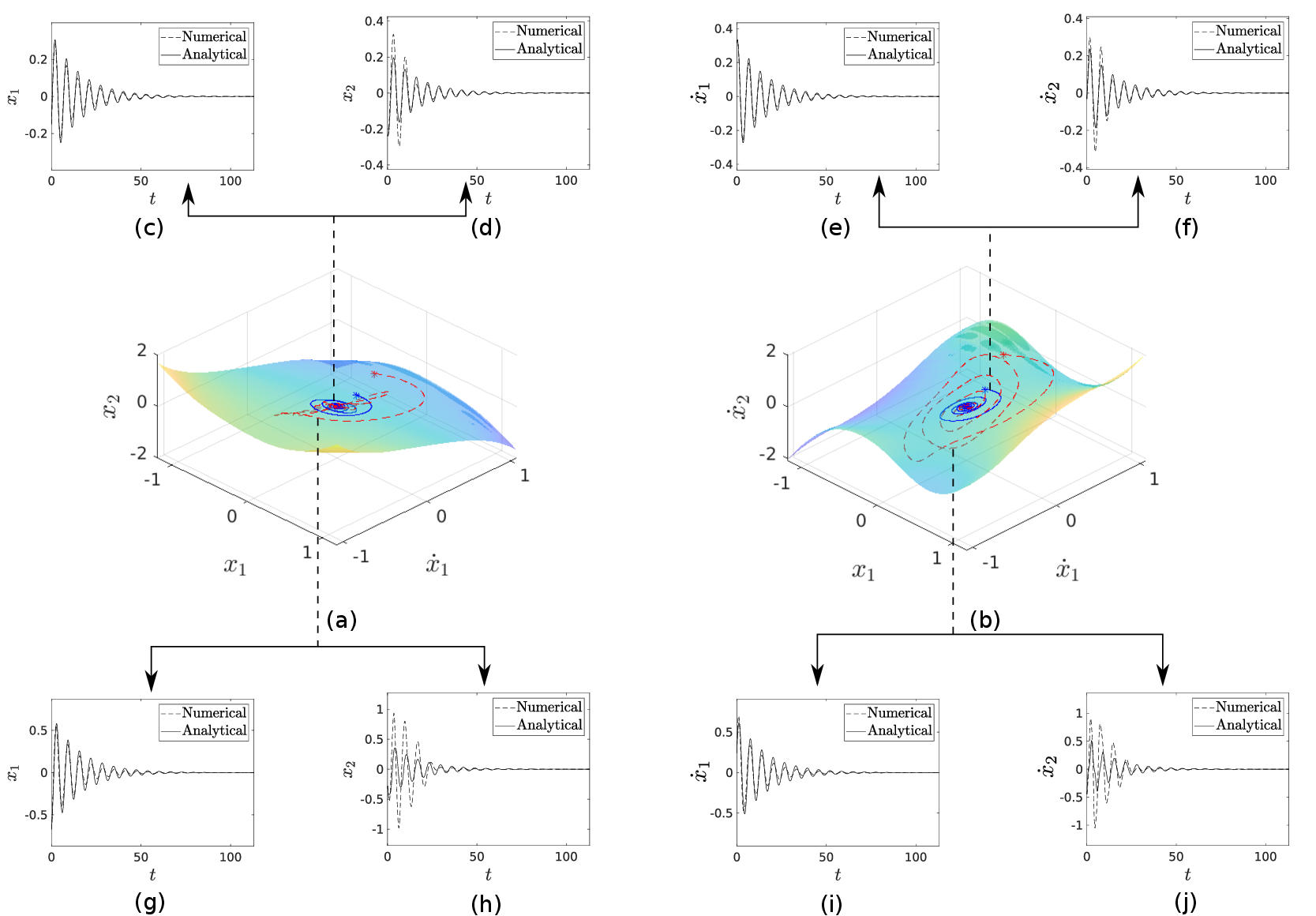}
\caption{NNM constructed with the Koopman modes: The projections of the invariant manifold on (a) the $x_1\dot{x}_1x_2$ space and (b) the $x_1\dot{x}_1\dot{x}_2$ space, are shown with the trajectories from numerical data. The time histories of all state variables: (c)-(f) correspond to the lower energy trajectories and (g)-(j) correspond to the higher energy trajectories.}\label{fignum6}
\end{figure}

In pursuit of the final objective of this study, the NNM manifolds are presented next, 
along with two trajectories that initiate on them - one with higher energy and another with lower energy; see Figs.\ref{fignum5} -\ref{fignum6}.  
The corresponding time histories are also presented, facilitating a comparison between analytical and numerical results. The NNM manifold, depicted as the  surface in Figs.\ref{fignum5}(a) \& (b), is calculated for the parameter set $\{ \varepsilon, \delta, \mu, \alpha, \beta \}=\{0.15,0.31,1,2,3 \}$. The trajectory represented by the blue curve on the manifold corresponds to the initial point having lower energy, whereas the trajectory shown in red (dashed curve) corresponds to the initial point with higher energy. A comparison of the time histories obtained analytically using the proposed framework 
along with those obtained numerically  is presented 
to show the accuracy of the approximations of $x_1, x_2, \dot{x}_1 \ {\rm and} \ \dot{x}_2$ respectively for the lower energy trajectory (the blue curve), in Figs. \ref{fignum5}(c)-(f). In Figs. \ref{fignum5}(g)-(j), the same is illustrated for the higher energy trajectory (the red dashed curve). It is observed that the approximation for the lower energy trajectory  is quite precise. However, the  accuracy slightly decreases in the comparison between $x_2$ and $\dot{x}_2$ for  the higher energy trajectory 
although the prediction is decent enough for practical purposes.
 In this case, the linear part of the stiffness of the second oscillator is $\varepsilon \cdot \delta=0.0465$, which is $\mathcal{O}(10^{-2})$, comparatively much smaller than the stiffness of the primary oscillator, which is $\omega_0=1$. Despite being closely related to the essentially nonlinear oscillator (ENO), this system exhibits a high level of accuracy in approximation. 
 

Sacrificing some accuracy allows for the analysis of a system even closer to the ENO, as demonstrated for the case  shown in Fig.\ref{fignum6}.
Here the parameters considered for the numerical example are taken to be $\{ \varepsilon, \delta, \mu, $ $\alpha, \beta \}=\{0.1,0.26,0.6,9,4 \}$, which makes $\varepsilon \cdot \delta=0.026$. As can be observed in Fig.\ref{fignum6}, 
the approximation of the lower-energy trajectory also seems to be  affected, and the analytical predictions of $x_2$ and $\dot{x}_2$ are not completely accurate. Furthermore, it is observed that the higher-energy trajectory is moving out of the manifold because of the comparatively lower accuracy of the approximation. The notable decrease in accuracy can be solely attributed to the arbitrary reduction of the linear stiffness component of the secondary oscillator, which causes the system to approach a state closer to that of an essentially nonlinear system and leads to  at least one of the eigenvalues approach zero.
As discussed in Section \ref{sec4}, this leads to  violating the first criterion of the Sternberg theorem (Theorem \ref{thm3}), necessitating all the fixed points being hyperbolic. Consequently, when an eigenvalue approaches zero, the determinant of the Jacobian matrix tends towards zero, {\it i.e.}, $\det(\bold{A}) \rightarrow 0$. As $\bold{A}^{-1}$ needs to be evaluated at each step of the computation of Koopman modes, a determinant close to zero leads to accumulation of errors compromising the accuracy of the Koopman operator-based framework.


\section{Conclusions}\label{sec8}
This study focused on the applicability and robustness of the Koopman operator framework across a broad parameter regime. The key highlights of this investigation are as follows:
\begin{enumerate}
    \item[(a)] The Proposition \ref{pro1}, which enables unique mapping of the state variables on to the space of observables,  is necessary for the further advancement of the procedure.  While this was used in \cite{cirillo2016spectral}, the uniqueness of the mapping is proved here for the first time. 
    \item[(b)] A connection of the Koopman eigenfunction with the diffeomorphism stated in the Hartman theorem (Theorem \ref{thm2}) has also been established using the theorems on linearization and from Corollary \ref{corl1}.
    
    \item[(c)] Further, it has been established the importance of 
    Sternberg's theorem (Theorem \ref{thm3}), which ensures the existence of the analytic diffeomorphism of the desired invariant manifold in the neighborhood of the hyperbolic fixed point, in the context of 
    the nonlinear state space transformation used in the proposed framework. 
    \item[(d)] The effects of damping and internal resonance on the Koopman modes have been studied through  extensive parametric  numerical investigations. 
    As damping contributes to the real part of the eigenvalue whereas the instantaneous frequency contributes to the imaginary part of the eigenvalue, in some sense this is somewhat equivalent to observing the Koopman modes at different regimes of eigenvalues. 
    \item[(e)] The limitations of the proposed framework for systems with characteristics similar to ENO have been  analysed.  
\end{enumerate}
It emerges from this study 
that the accuracy of the Koopman operator framework strongly depends on the damping of the system and the closeness of the system to internal resonance. 
According to Sternberg theorem, the existence of resonating eigenvalues forfeits the analytic diffeomorphism of the manifold which otherwise naturally exists in the neighbourhood of the origin, leading towards folding of the manifold. In this paper, such cases 
of resonating eigenvalues have been avoided  but the presence of strong nonlinearity ensures nonlinear internal resonance which appears to be beneficial for the proposed framework. Although this framework fails for systems with ENO,  the framework is  shown to be quite robust for a system that is very similar to ENO. 





\section*{Declarations}
\subsection*{Authors' contributions:} {\bf Rahul Das:} Conceptualization, Formulation, Methodology, Visualization, Software, Writing – original draft. {\bf Anil K. Bajaj:} Supervision, Writing – review \& editing. {\bf Sayan Gupta:} Supervision, Project management, Writing – review, editing \& final draft writing.
\subsection*{Funding:} This study is funded by Ministry of Education, Government of India, under the Institute of Eminence scheme for the Center for Complex Systems \& Dynamics, IIT Madras with  Project No. SP2021077/ DRMHRD/ DIRIIT.
\subsection*{Conflict of interest:} The authors declare that they have no known competing financial interests or personal relationships that could have appeared to influence the work reported in this paper.







\begin{appendices}
\section{Modal lines, similar and non-similar NNMs}\label{secA}
The fundamental concept behind defining normal modes is to establish the functional relationship between degrees of freedom, leading to their visualization in the configuration space formed by these degrees of freedom. Importantly, LNMs naturally involve a linear relationship, making their representation on the configuration plane as straight lines.
\begin{figure}
    \centering
    \includegraphics[width=0.45\textwidth]{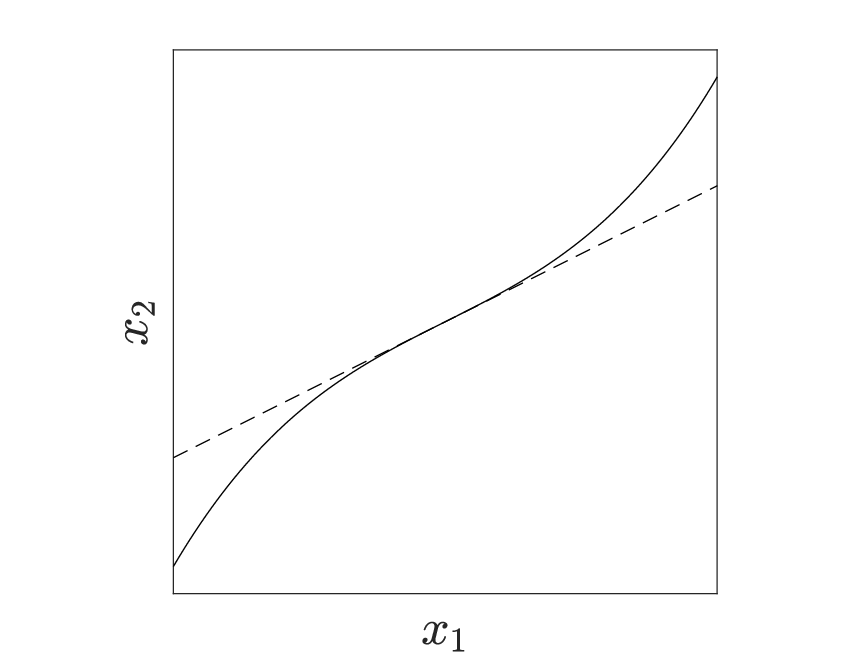}
    \caption{Schematic representation of modal lines on the configuration plane $x_1x_2$. The dashed modal line corresponds to a similar NNM motion, whereas the modal curve drawn in solid line corresponds to a non-similar NNM motion.}
    \label{Modallines}
\end{figure}
\begin{enumerate}
    \item[(a)] {\bf Modal lines:} These curves on the configuration space, representing the relationship between degrees of freedom, are called modal lines. For linear dynamical systems, modal lines are always straight lines, with the functional dependence expressed as $x_2=c \cdot x_1$; where $x_1$ and $x_2$ are the degrees of freedom of the system, and $c \in \mathbb{R}$ is a scalar coefficient. The schematic representation in Fig.\ref{Modallines} illustrates two different classes of modal lines. 
    \item[(b)] {\bf Similar NNMs:}  While the functional dependence in nonlinear systems is typically nonlinear, special cases like homogeneous and symmetric systems exhibit linear dependencies between degrees of freedom, expressed as, $x_2=c \cdot x_1$. Correspondingly, modal lines remain straight, resembling those in linear systems. These representations, termed similar NNMs, are illustrated by the dashed line in Fig.\ref{Modallines}, where the slope of the line represents the scalar $c$. Their behavior closely mirrors that of linear systems, simplifying their quantification.
    \item[(c)] {\bf Non-similar NNMs:} In scenarios where the functional dependence is nonlinear, excluding special cases (homogeneous and symmetric systems), $x_2$  becomes some nonlinear function of $x_1$, denoted as, $x_2=\widetilde{x}_2(x_1)$, where $\widetilde{x}_2(\cdot)$ is a nonlinear function. The corresponding modal lines manifest as curves on the configuration space. In Fig.\ref{Modallines}, the schematic representation of the more general scenario of nonlinear dynamical systems, \textit{i.e.}, the non-similar NNMs is shown with the solid curve. These non-similar NNMs pose greater challenges in quantification due to their nonlinear functional relations between degrees of freedom.
\end{enumerate}
\vspace{0.3cm}
\section{Orthogonality of eigenvectors}\label{secB}
Consider a $n \times n$ diagonalizable matrix $\bold{A}$ with $n$ distinct eigenvalues $\lambda_i$, where $i \in \{ 1, \ldots, n\}$. The eigenvalues of the adjoint matrix $\bold{A}^*$ are denoted $\Bar{\lambda}_i$. Consider an arbitrary eigenvalue $\lambda_i$ and the set of corresponding eigenvectors $\Phi_i$ (as the eigenvector is not unique). Similarly, let the eigenvalue  for  $\bold{A}^*$ be $\Bar{\lambda}_j$ and the set of corresponding eigenvectors be $\Phi_j$. Choose a vector $\alpha \in \Phi_i$ and another vector $\beta \in \Phi_j$. As these  obviously are the eigenvectors corresponding to $\lambda_i \ {\rm and} \ \Bar{\lambda}_j$ respectively, one can write
\begin{equation}
\label{appendixA1}
    \centering
    \bold{A}\alpha=\lambda_i \alpha, \ {\rm and} \ \bold{A}^*\beta=\Bar{\lambda}_j \beta.
\end{equation}
Now,
\begin{align}
	\label{appendixA2}
	(\lambda_i - \lambda_j) \langle \alpha | \beta \rangle & = \lambda_i \langle \alpha | \beta \rangle - \lambda_j \langle \alpha | \beta \rangle \notag \\
	& = \langle \lambda_i \alpha | \beta \rangle - \langle \alpha | \Bar{\lambda}_j \beta \rangle \quad \text{(using properties of inner product \cite{hoffmann1971linear})} \notag \\
	& = \langle \mathbf{A} \alpha | \beta \rangle - \langle \alpha | \mathbf{A}^* \beta \rangle \quad \text{(using Eq.\eqref{appendixA1})}.
\end{align}
From the properties of inner product,  we know that 
For any two vectors $X \ {\rm and} \ Y$ of same dimension, one gets
\begin{equation}
    \label{appendixA3}
    \centering
    \langle X | Y \rangle = \sum_i X_i \Bar{Y}_i = Y^*X,
\end{equation}
which leads to
\begin{equation}
    \label{appendixA4}
    \centering
    \langle \bold{A} X | Y \rangle = Y^* \bold{A}X=(\bold{A}^*Y)^*X=\langle X | \bold{A}^*Y \rangle.
\end{equation}
From Eq.\eqref{appendixA2} and Eq.\eqref{appendixA4} it follows that 
\begin{equation}
(\lambda_i-\lambda_j) \langle \alpha | \beta \rangle=0,
\end{equation}
which implies either $(\lambda_i-\lambda_j) =0$ or $\langle \alpha | \beta \rangle=0$. Since our choice of $\bold{A}$ ensures that there is no multiplicity of the eigenvalues, \textit{i.e.}, $\lambda_i \neq \lambda_j$ if $i \neq j$, therefore $\langle \alpha | \beta \rangle=0$.
We normalize the eigenvectors of $\bold{A}$ corresponding to eigenvalue $\lambda_i$ for uniqueness and represent them as, $v_i \in \Phi_i$ (which can be visualized as $v_i=c_i\alpha; \ c_i \in \mathbb{C}$); likewise the eigenvectors of $\bold{A}^*$ corresponding to eigenvalue $\lambda_j$ can be represented as $w_j \in \Phi_j$ (which can be visualized as $w_j=c_j\beta; \ c_j \in \mathbb{C}$).
Thus the orthonormal property of the eigenvectors can be represented as
\begin{equation}
    \centering
    \langle v_i | w_j \rangle =\begin{cases}
    1, & i=j \\
    0, & i \neq j.
    \label{appendixA5}
    \end{cases}
\end{equation}
\section{Theorems on the linearization of a nonlinear vector field} \label{sec_def}
For convenience, the theorems with some definitions are stated as follows,
\vspace{0.3cm}
\begin{definition}[Homeomorphism] Given two topological spaces $A$ and $B$, a continuous map $h \colon A \rightarrow B$ is called homeomorphism iff $h$ is a bijection and its inverse, $h^{-1}$ is continuous as well.
 \end{definition}
\vspace{0.3cm}

\begin{definition}[Manifold] Let $M$ be a topological space. For any arbitrary $m \in M$, iff  $\exists$ a homeomorphism between $M$ and Euclidean space, i.e. $h \colon M \rightarrow \mathbb{R}^n$, $M$ is called a manifold. More precisely a $n$-dimensional manifold.
 \end{definition}
\vspace{0.3cm}

\begin{definition}[Diffeomorphism] Given two manifolds $M$ and $N$, a differentiable map $d \colon M \rightarrow N$ is called a diffeomorphism iff it is a bijection and its inverse, $d^{-1} \colon N \rightarrow M $ is differentiable as well. If these functions are $r$ times continuously differentiable, $d$ is called a $C^r$- diffeomorphism.
 \end{definition}
\vspace{0.3cm}
\begin{theorem}[Hartman-Grobman theorem]\label{thm1}
 Let $\bold{f} \in C^1(U)$. Suppose that $\bold{A}$ has no eigenvalues with zero real part. Then $\exists$ a homeomorphism $\bold{h}$ of an open set $\mathcal{U} \subset U, \bold{0} \subset \mathcal{U}$ onto an open set $\mathcal{V} \subset \mathbb{R}^n, \bold{0} \in \mathcal{V}$ such that for each $\bold{x}_0 \in \mathcal{U}, \exists$ an open interval $\mathcal{I} \subset \mathbb{R}$ containing zero such that $\forall \ \bold{x}_0 \in \mathcal{U}$ and $t \in \mathcal{I}$;
 \begin{equation}
     \centering
     \bold{h} \circ \boldsymbol{\phi}(\bold{x},t)=\exp{(\bold{A}t)}\bold{h}(\bold{x}_0);
     \label{24}
 \end{equation}
 \textit{i.e.}, \rm{ $\bold{h}$ maps the trajectories of Eq. \eqref{21} near the origin to the trajectories of the linear system given by Eq. \eqref{27} and preserves the parametrization by time}.
  \end{theorem}

\vspace{0.3cm}

\section{Tangent Space}\label{secD}

\begin{figure}[htbp]%
\centering
\includegraphics[width=0.9\textwidth]{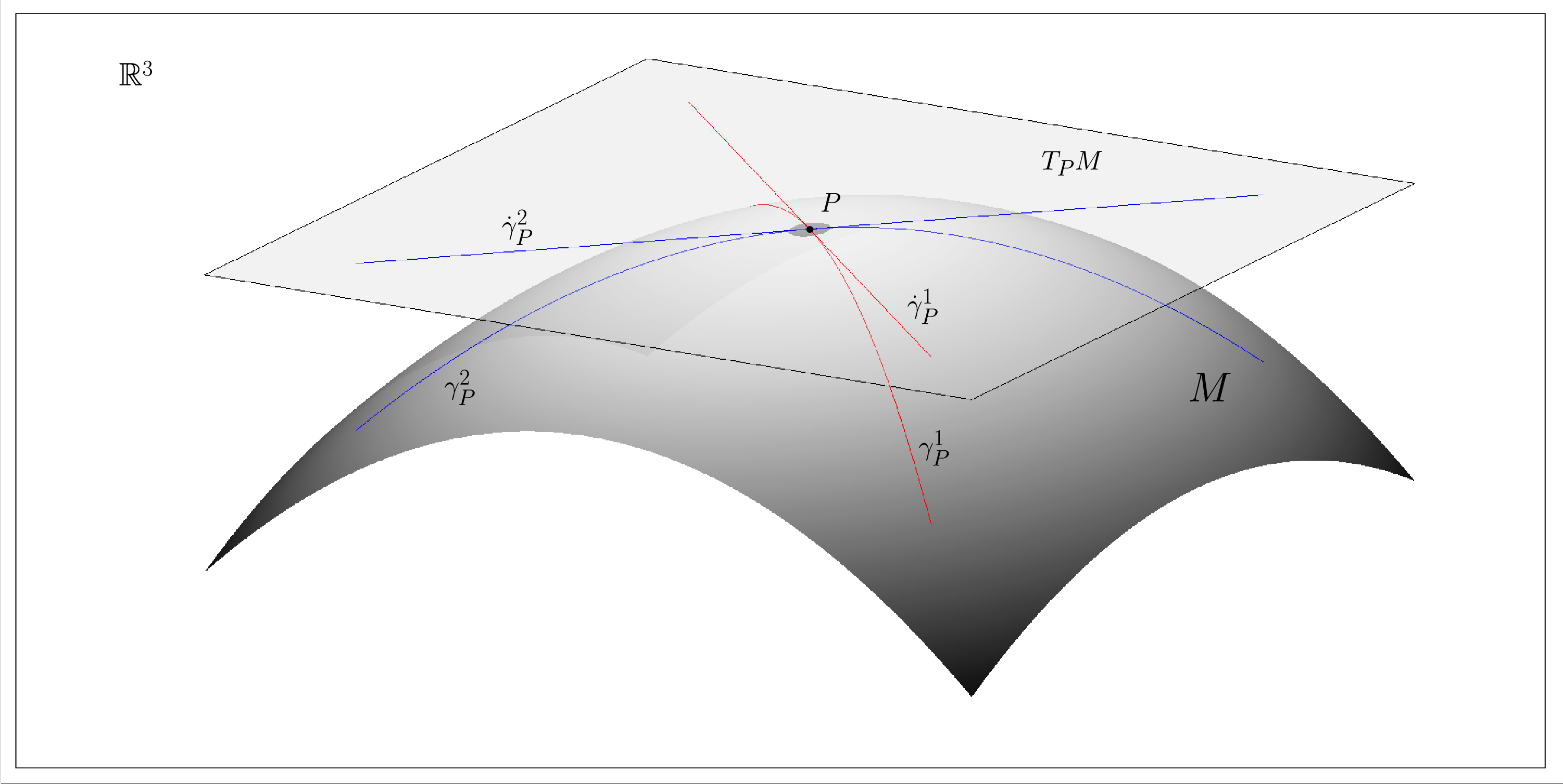}
\caption{$T_PM$ is the tangent space of the $2$-dimensional manifold $M$, embedded in $\mathbb{R}^3$ (for visualization we kept it restricted in $3$-dimensional space, which can be generalized up to any $\mathbb{R}^n$), at the point $P$. The vectors $\dot{\gamma}_P^1$ and $\dot{\gamma}_P^2$ are tangents of the differentiable function $\gamma_P^1$ and $\gamma_P^2$ respectively at the point $P$, which spans the tangent space $T_PM$.}\label{figappendixB1}
\end{figure}

\paragraph{Tangent bundle:}
In differential geometry, the \textit{tangent bundle} of a differentiable manifold $M \subseteq \mathbb{R}^n$, is a manifold $T M$ which assembles all the tangent vectors at every point on $M$ \cite{schlichtkrull2014differentiale}. As a set, it is given by $T M \subseteq M \times \mathbb{R}^n$, and it is defined as,
\begin{equation}
    \label{appendixB1}
    \centering
    T M := \{ (P,\dot{\gamma}_P) \ | \ \gamma_P \colon (-\varepsilon, \varepsilon) \rightarrow M \},
\end{equation}
where $\gamma_P(0)=P \in M$, represents a point on the manifold $M$ and $\dot{\gamma}_P=\frac{d }{d t}(\gamma_P(t)) \in \mathbb{R}^n$ is a tangent of the smooth curve $\gamma_P(t)$, at the point $P$.
The addition between two elements of the tangent bundle $T M$ is defined as $\dagger \colon T M \rightarrow T M$, such that,  
\begin{equation}
    \label{appendixB2}
    \centering
    (P,v) \dagger (P,w) = (P,v+w); \ v,w \in \mathbb{R}^n,
\end{equation}
where the map $+ \colon \mathbb{R}^n \rightarrow \mathbb{R}^n$ denotes the addition between two elements of $\mathbb{R}^n$. 
\paragraph{Tangent space of a manifold:}
Tangent space of a manifold $M$ at a point $P$, denoted with $T_PM$ is a proper subset of the tangent bundle $T M$, such that the tangent bundle is the disjoint set union of all the tangent spaces of the manifold $M$, defined by,
\begin{equation}
    \label{appendixB3}
    \centering
    T M :=\bigsqcup_{P \in M} T_PM= \bigcup_{P \in M} \{ P \} \times T_PM=\bigcup_{P \in M} \{ (P,q) \ | \ q \in T_PM \}.
\end{equation}
So from the the definition given in Eq. \eqref{appendixB1} and Eq. \eqref{appendixB3} we can conclude that $q=\dot{\gamma}_P \in T_PM \subset \mathbb{R}^n$, the schematic of the tangent space of the manifold is represented in Fig.\ref{figappendixB1}. Now our objective is to determine the equation of the tangent space of a manifold $M$ at a point $P$, for that, let us consider a differentiable function, defined as $f \colon \mathbb{R}^n \rightarrow \mathbb{R}$, and let a manifold $M$ is constructed by the zero level set of the function $f$, such that,
\begin{equation}
    \label{appendixB4}
    \centering
    f(m)=0; \ \forall \ m \in M.
\end{equation}
We want to define $T_PM$ in terms of $f$. Now from the definition $\gamma_p(t) \in M$, and according to Eq. \eqref{appendixB4},
\begin{align}
    \label{appendixB5}
    \centering
    & f \circ \gamma_P(t)  = 0 \notag \\
   \implies & \frac{d}{d t}(f \circ \gamma_P(t)) = 0 \notag \\
   \implies & \nabla f |_p \cdot \dot{\gamma_P} |_{t=0} =0 \notag \\
   \implies & \begin{pmatrix}
       \frac{\partial f}{\partial x_1} \ \frac{\partial f}{\partial x_2} \ \cdots \frac{\partial f}{\partial x_n}
   \end{pmatrix}|_P \cdot \begin{pmatrix}
       \dot{\gamma}_1|_{t=0} \\ \dot{\gamma}_2|_{t=0} \\ \vdots \\ \dot{\gamma}_n|_{t=0}
   \end{pmatrix}=0.
\end{align}
Eq. \eqref{appendixB5} represents a general form of the equation of tangent space of manifold $M$ at a point $P$. For our case described in Section \ref{sec5}, we require the equation of tangent space at the origin. For that without loss of generality we can fix the origin at $P$ and set our coordinate axes along $\dot{\gamma}_i|_{t=0}; \ i=1,2,\ldots,n$, such that,
\begin{equation}
    \label{appendixB6}
    \centering
    \bold{x}=\dot{\gamma}_P|_{t=0},
\end{equation}
and the equation of the tangent space will become,
\begin{equation}
    \label{appendixB7}
    \centering
    \frac{\partial f}{\partial \bold{x}} \cdot \bold{x}=0,
\end{equation}
where $\frac{\partial f}{\partial \bold{x}}$ represents the Jacobian of the function $f$. Eq. \eqref{appendixB7} generalizes the concept of tangent of a function $f$ (passing through origin), at origin, which is $m \cdot x =0$, where $m=\frac{d f}{d x}|_{x=0}$, is the slope of $f$ at the origin. This is an example of an one dimensional tangent space of the function $f$.
\end{appendices}






\bibliography{bibliography}

\end{document}